\documentclass[vecphys]{svmult}
\usepackage{makeidx}         
\usepackage{graphicx}      
\usepackage{multicol}     
\usepackage[bottom]{footmisc}
\makeindex                                                
\newcommand{\beq}{\begin{equation}}
\newcommand{\eeq}{\end{equation}}
\newcommand{\bea}{\begin{eqnarray}}
\newcommand{\eea}{\end{eqnarray}}

\def\laq{~\raise 0.4ex\hbox{$<$}\kern 
-0.8em\lower 0.62ex\hbox{$\sim$}~}
\def\gaq{~\raise 0.4ex\hbox{$>$}\kern 
-0.7em\lower 0.62ex\hbox{$\sim$}~}

\def \na {\nabla}
\def \pa {\partial}
\def \ti {\widetilde}
\def \wh {\widehat}
\def \pr {{\prime}}

\def \ra {\rightarrow}

\def \La {\Lambda}
\def \da {\delta}
\def \Da {\Delta}
\def \b {\beta}
\def \a {\alpha}
\def \ap {\alpha^{\prime}}
\def \Ga {\Gamma}
\def \ga {\gamma}
\def \sg {\sigma}
\def \Sg {\Sigma}
\def \ep {\epsilon}

\def \r {\rho}
\def \om {\omega}
\def \Om {\Omega}
\def\vp{\varphi}

\def \ls {\lambda_{\rm s}}
\def \lp {\lambda_{\rm P}}
\def \Ms {M_{\rm s}}
\def \Mp {M_{\rm P}}

\def \fpu {\dot{\phi}}
\def \fpp {\ddot{\phi}}
\def \fb {\overline \phi}
\def \fbp {\dot{\fb}}
\def \fbpp {\ddot{\fb}}
\def \rb {\overline \rho}
\def \pb {\overline p}
\def \sgb {\overline \sg}
\def \ds {{\delta \sigma}}
\def \hf {\widehat \phi}
\def \hfd {{\dot{\hf}}}

\def \rg {\sqrt{-g}}
\def \rp {\sqrt{\ep (\nabla \phi)^2}}
\def \cl {{\cal L}_m}

\def \d {{\rm d}}
\def \e {{\rm e}}

\begin{document}

\thispagestyle{empty}

\begin{flushright}
BA-TH/06-558\\
hep-th/0702166
\end{flushright}

\vspace{1.5 cm}

\begin{center}

\huge{Dilaton Cosmology \\and Phenomenology}

\vspace{1cm}

\large{M. Gasperini}

\bigskip
\normalsize

\smallskip
{\sl Dipartimento di Fisica, Universit\`a di Bari, \\
Via G. Amendola 173, 70126 Bari, Italy}

and 

\smallskip
{\sl Istituto Nazionale di Fisica Nucleare, Sezione di Bari, 
Bari, Italy}

\smallskip
\smallskip
E-mail: {\tt gasperini@ba.infn.it}

\vspace{1cm}

\begin{abstract}
This paper is dedicated to Gabriele Veneziano on his  $65th$ birthday. Most of the results reported here are known results, due to Gabriele, or obtained in collaboration with him, or inspired by our joint work on string cosmology. A few new results are also presented concerning the duality invariance of a  non-local dilaton coupling to the matter sources, and its possible cosmological applications in the context of the dark-energy scenario. 
\end{abstract}
\end{center}

\vspace{0.5cm}
\begin{center}
------------------------------------

\vspace{0.5cm}
Contribution to the book: \\
\bigskip
{\em ``String theory and fundamental interactions:}\\
\smallskip
{\em celebrating Gabriele Veneziano on his 65th birthday"},\\
\smallskip
eds. M. Gasperini and J. Maharana\\
\smallskip
(Lecture Notes in Physics, Springer-Verlag, Berlin/Heidelberg, 2007)\\ 
\bigskip{\tt www.springerlink.com/content/1616-6361}
\end{center}
\newpage

\title*{Dilaton cosmology and phenomenology}
\author{M. Gasperini}
\authorrunning{M. Gasperini} 
\institute{Dipartimento di Fisica , Universit\`a di Bari, 
Via G. Amendola 173, 70126 Bari, Italy, and Istituto Nazionale di Fisica Nucleare, Sezione di Bari, 
Bari, Italy, 
\texttt{gasperini@ba.infn.it}}

\maketitle

\begin{abstract}
This paper is dedicated to Gabriele Veneziano on his  $65th$ birthday. Most of the results reported here are known results, due to Gabriele, or obtained in collaboration with him, or inspired by our joint work on string cosmology. A few new results are also presented concerning the duality invariance of a  non-local dilaton coupling to the matter sources, and its possible cosmological applications in the context of the dark-energy scenario. 
\end{abstract}

\section*{Foreword and introduction}
\label{sec:1}

My collaboration with Gabriele Veneziano has continued, almost uninterruptedly, for more than fifteen years (even now we are preparing a joint contribution to the book {\em ``Beyond the Big Bang"}, which will be published by Springer, as this book). Our first meeting dates back to 1989, when Gabriele came to the University of Turin to give a series of talks and seminars. At that time I was working there as a young researcher at the Department of Theoretical Physics, and I remember that Sergio Fubini (professor at the same Department)  introduced me to Gabriele before a seminar. After the seminar we went to my office, and we started talking about cosmology, big bang, inflation, and strings. Gabriele was able to make me feel at ease, in spite of the fact that I was a bit embarrassed, being face to face with such a world-renown scientist like him: before that meeting, indeed, I knew him  only for having seen his name quoted in many books and articles as one of the founders of string theory. I could not imagine that I was about to embark on the most stimulating and important adventure of my scientific life. 

\begin{figure}
\centering
\includegraphics[height=7.5cm]{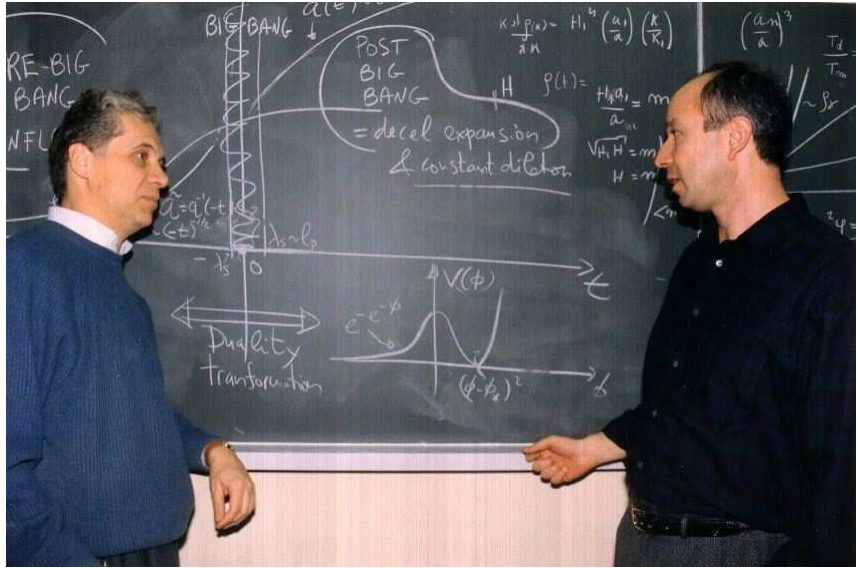}
\caption{Gabriele Veneziano (left) and the author (right), talking about dilatons at CERN (January 1994).}
\label{Veneziano&Gasperini}      
\end{figure}

After that meeting we started collaborating on string cosmology, and I visited very often the Theory Division (now ``TH Unit") at CERN, living in Geneva also for long periods. This has given me the opportunity to appreciate Gabriele not only as a scientist -- whose inventiveness, originality, profundity of thought will not be stressed here, because they are well-known to the physicist community -- but also for his human qualities. His tireless enthusiasm for physics, his generosity in sharing knowledge, his intellectual honesty, always make working with him a rewarding and enjoyable experience.  I have countless memories of days spent discussing and working out calculations on the blackboard of his office (see Fig. \ref{Veneziano&Gasperini}), with short ``coffee breaks" every now and them, talking about physics even during lunch and dinner. Countless are the things I have learned from him, not only from a scientific but also from a human point of view. I will be grateful to him forever. 

Choosing among the lines of research developed in collaboration with Gabriele, I will concentrate the contribution to this book on the possible role played by the dilaton in a cosmological context, with particular attention to the phenomenological aspects of dilaton cosmology. The dilaton is a fundamental scalar field appearing in all models of superstrings, dilaton cosmology is probably the most natural and typical form of ``string cosmology", and a direct/indirect confirmation (or disproof) of its predictions could give us important experimental information on string theory in general. 

The present contribution contains three lectures. The first lecture (Sect. \ref{sec1}) is devoted to the presentation of a primordial cosmological scenario in which the background evolution is dominated by the dilaton, and the Universe is driven through an accelerated phase representing the ``dual" counterpart of the standard, decelerated evolution. The second lecture (Sect. \ref{sec2}) will discuss the possibility that a cosmic background of relic dilaton radiation could have survived until present, and could be detectable by the gravitational antennas that are presently operating (or planned for a near-future operation).  Finally, the third lecture (Sect. \ref{sec3}) will suggest a possible ``dilatonic" origin of the dark-energy fluid dominating the cosmic acceleration recently observed on large scales, stressing the main differences from other, more conventional models of scalar ``quintessence". 

\subsubsection{Notations and conventions}

Unless otherwise stated, the following conventions are used throughout this paper: Greek indices run from $0$ to $d$, Latin indices from $1$ to $d$, where $d=D-1$ is the number of spatial dimensions of the $D$-dimensional space-time manifold. The metric signature is: 
\bea
&&
g_{\mu\nu} = {\rm diag} (+, -, -,-, \cdots).
\nonumber 
\eea
The Riemann tensor and its contractions are defined by:
\bea
&&
R_{\mu\nu\a}\,^\b=\pa_\mu \Ga_{\nu\a}\,^\b+ 
\Ga_{\mu\rho}\,^\b \Ga_{\nu\a}\,^\rho - (\mu \leftrightarrow \nu),
\nonumber \\ &&
R_{\nu\a}= R_{\mu\nu\a}\,^\mu, ~~~~~~~~~~~~
R= R_\mu\,^\mu, ~~~~~~~~~~~~
G_{\mu\nu}=R_{\mu\nu}-{1\over 2} g_{\mu\nu} R.
\nonumber
\eea
The conventions for the covariant derivative are:
\bea
&&
\nabla_\mu V_\a= \pa_\mu V_\a -\Ga_{\mu\a}\,^\b V_\b,
~~~~~~~~~\nabla_\mu V^\a= \pa_\mu V^\a +\Ga_{\mu\b}\,^\a V^\b.
\nonumber 
\eea
Finally, we often use the convenient notation:
\bea
&&
(\nabla \phi)^2\equiv \nabla_\mu \phi \nabla^\mu \phi, ~~~~~~~~~~~~
\nabla^2 \phi \equiv \nabla_\mu  \nabla^\mu \phi.
\nonumber 
\eea

\section{Dilaton-dominated inflation: the pre-big bang scenario}
\label{sec1}

If we apply to the ``specialized" literature for a description of the birth and of the first moments of our Universe, we may read, in the (probably) most ancient and  authoritative  book, that

\begin{itemize}
\item[]
{\em ``In the beginning God created the Heaven and the Earth, \\
and the Earth was without form, and void;\\
and the darkness was upon the face of the deep.\\
And the Breath of God\\
moved upon the face of the water."}
\begin{flushright}
(Genesis, The Holy Bible).
\end{flushright}
\end{itemize}
The most impressive aspect of these verses, for a modern cosmologist, is probably the total absence of any reference to the  extremely hot, kinetic, explosive state that one could expect at (or immediately after) the ``big bang" deflagration. The described state, instead, is somewhat quiet, dark, empty -- we can read, indeed,  about ``void", ``darkness", and ``the deep" gives us the idea of something enormously desert and empty. In this static configuration there is at most some small fluctuation (the ``Breath"), a ripple on the surface of this vacuum.

It is amusing to note that a state of this type (flat, cold and vacuum, only ruffled by quantum fluctuations), can be obtained as the initial state of our Universe, in a string-cosmology context, under the hypothesis that the Universe evolves in a ``self-dual" way with respect to the symmetries of the low-energy string effective action \cite{1,2}.

To introduce this result we start considering the gravi-dilaton sector of the low-energy effective action. To lowest order in the $\ap$ (higher-derivative) and $g_s^2$ (higher-loop) expansion the action is the same for all models of superstrings \cite{3,4}, and is given by 
\beq
S=-{1\over 2\ls^{d-1}} \int_{\Om} \d^{d+1} x \sqrt{-g} \, \e^{-\phi} \left(R+ \pa_\mu\phi\,\pa^\mu \phi +V\right).
\label{1}
\eeq
Here $\phi$ is the dilaton, and $\ls=(2\pi \ap)^{1/2}$ is the fundamental 
length parameter of string theory. We have written the action using the so-called String frame (S-frame) metric, i.e. the metric to which a ``test" string is minimally coupled and in which its evolution is geodesics. We have also included, for completeness and for further applications,  a (possibly non-perturbative) dilaton potential, $V=V(\phi)$. 

This action should be completed by the source term $S_m(g, \phi)$, describing the matter-fields contributions, and by the Gibbons-Hawking boundary term $S_\Sg$, which is required (as in general relativity) to cancel the variational contributions of the second derivatives of the metric following from the Einstein-Hilbert Lagrangian $\sqrt{-g}\,R$. For the S-frame action (\ref{1}) the boundary term takes the form \cite{5}
\beq
S_\Sg= {1\over 2\ls^{d-1}} \int_{\pa \Om} \sqrt{-g}~\e^{-\phi} K^\a
\d \Sg_\a,
\label{2}
\eeq
where $K^\a= K n^\a$. Here $K$ is the trace of the extrinsic curvature of the $d$ dimensional hypersurface $\pa \Om$ bounding the 
hypervolume $\Om$ over which we are varying the action, and $n^\a$ is the unit  vector normal to this hypersurface. 

The variation of the total action $S+S_m+S_\Sg$ with respect to $g^{\mu\nu}$ leads then to the equations 
\beq
G_{\mu\nu}+\nabla_\mu \nabla_\nu \phi  + {1\over 2} g_{\mu\nu} \left(\nabla \phi\right)^2
-g_{\mu\nu} \nabla^2 \phi- {1\over 2} g_{\mu\nu}V(\phi)
=\ls^{d-1}\e^\phi ~T_{\mu\nu},
\label{3}
\eeq
where $G_{\mu\nu}$ is the Einstein tensor and $T_{\mu\nu}$ the gravitational stress tensor of the matter sources, defined as usual by the functional differentiation of $S_m$ as:
\beq
T_{\mu\nu}= {2\over \sqrt{-g}} {\da S_m\over \da g_{\mu\nu}}.
\label{4}
\eeq
The variation of the total action  with respect to $\phi$ leads to the dilaton equation of motion,
\beq
R+ 2 \nabla^2 \phi-(\nabla \phi)^2+V- {\pa V\over \pa \phi}= \ls^{d-1} \e^{\phi} \sg, 
\label{5}
\eeq
where $\sg$ is the (S-frame) density of dilaton charge of the sources, defined by the functional differentiation of $S_m$ with respect to $\phi$:
\beq
\sg= -{2\over \sqrt{-g}} {\da S_m\over \da \phi}.
\label{6}
\eeq
Using the dilaton equation to eliminate the scalar curvature, present in the Einstein tensor, we can eventually rewrite Eq. (\ref{3}) in the convenient (simplified) form
\beq
R_{\mu\nu}+\nabla_\mu \nabla_\nu\phi - {1\over 2} g_{\mu\nu}{\pa 
V\over \pa \phi} = \ls^{d-1} \e^{\phi}\left(T_{\mu\nu}+{1\over 2} g_{\mu\nu}\, \sg\right).
\label{7}
\eeq

\subsection{Scale factor duality}
\label{sec11}

We will now consider the particular case in which the space-time manifold described by the S-frame metric is spatially flat, homogeneous (but not necessarily isotropic), and in which the matter fields  can be phenomenologically described as perfect fluids, at rest in the comoving frame of the given Robertson-Walker geometry. We can thus set, in the synchronous gauge, 
\bea
&&
g_{\mu\nu}= {\rm diag} (1,- a_i^2 \da_{ij}), ~~~~~~~ a_i=a_i(t),
~~~~~~~ \phi=\phi(t),
\nonumber\\ &&
T_\mu\,^\nu= {\rm diag} (\r, -p_i \da_i^j), ~~~~~~~~ \r=\r(t),
~~~~~~~~ p_i=p_i(t), ~~~~~~~ \sg = \sg (t).
\label{8}
\eea
Separating the time and space components of the gravitational equations we then obtain, from the $(00)$ component of Eq. (\ref{3}),
 \beq
 \fpu^2 -2  \fpu \sum_i H_i + \left( \sum_i H_i\right)^2- \sum_iH_i^2-V
 =2 \ls^{d-1}\,\e^\phi \r 
 \label{9}
 \eeq
(where $H_1= \dot a_i/a_i$). From the $(ii)$ component of Eq. (\ref{7}) we have:
 \beq
 \dot H_i -H_i \left(\fpu - \sum_k H_k\right) +{1\over 2} {\pa V\over \pa \phi} =   \ls^{d-1}\, \e^\phi \left( p_i- {\sg \over 2}\right).
 \label{10}
 \eeq
From the dilaton equation (\ref{5}) we are lead, finally, to
 \beq
2 \fpp - \fpu^2 +2 \fpu \sum_iH_i- \sum_i\left(2\dot H_i+H_i^2\right)
- \left( \sum_i H_i\right)^2 +V- {\pa V\over \pa \phi}=\ls^{d-1}\, \e^\phi \sg. 
 \label{11}
 \eeq
We have thus obtained a system of $d+2$ equations for the $2d+3$ unknowns $\{a_i, \phi, \r, p_i, \sg\}$: its solution requires the  input of $d+1$ ``equations of state", $p_i=p_i(\rho)$, $ \sg= \sg(\rho)$, specifying the properties of the considered matter sources. 

Let us now consider the symmetries of this system of equations.   
There are two symmetries, in particular, that are relevant for the discussion of this section. One of them (also present in the cosmological equations of general relativity) is the invariance under the time-reversal transformation $t \ra -t$, which implies
\beq
H_i \ra -H_i , ~~~~~~~~ \dot H_i \ra \dot H_i, ~~~~~~~~  \fpu \ra -\fpu, ~~~~~~~~ 
\fpp \ra \fpp.
\label{12}
\eeq
Thanks to this invariance property, if the set of variables $S=\{a_i(t), \phi(t), \r(t)\}$ represents an exact solution of Eqs. (\ref{9})-(\ref{11}), then the time-reversed set $\ti S=\{a_i(-t), \phi(-t), \r(-t)\}$ also corresponds to an exact solution of the same equations (with different kinematic properties, in general). 

The string-cosmology equations, in the particular case $\sg=0$ and $V=$ const,  are also invariant under other transformations which have no analogue in general relativity, and which include the inversion of an arbitrary number of scale factors of the background geometry (\ref{8}):  the so-called ``scale-factor duality" transformations \cite{1,6}. For a simple  illustration of this  property we may conveniently rewrite the equations in terms of the ``shifted variables" $\fb$, $\rb$, $\pb_i$, $\sgb$, defined by 
\bea
&&
\fb=\phi- \ln \prod_i a_i= \phi- \sum_i \ln a_i, 
~~~~~~~~~ i=1, \dots, d,
\nonumber\\ &&
\rb= \r \prod_i a_i, ~~~~~~~~~ \pb_k= p_k \prod_i a_i,
 ~~~~~~~~~ \sgb= \sg \prod_i a_i.
\label{13}
\eea
Eqs. (\ref{9})-(\ref{11}) then become:
\bea
&&
\fbp^2-\sum_i H_i^2-V=2 \ls^{d-1}\,\e^{\fb}~ \rb, 
\label{14}\\ &&
\dot H_i- H_i \fbp+{1\over 2} {\pa V\over \pa \phi}= \ls^{d-1}\,\e^{\fb} 
\left(\pb_i- {\sgb\over 2}\right),
\label{15}\\ &&
2 \fbpp - \fbp^2 -\sum_i H_i^2+V- {\pa V\over \pa \phi}=\ls^{d-1}\, \e^{\fb} ~ \sgb. 
\label{16}
\eea
Under the transformation $a \ra \ti a =a^{-1}$, on the other hand, we have:
\beq
H= a^{-1} {\d a \over \d t}~~ \ra ~~ \ti H =  \ti a^{-1} {\d \ti a \over \d t} = a {\d a^{-1}\over \d t}= -H.
\label{17}
\eeq
We can then easily check that Eqs. 
(\ref{14})-(\ref{16}), in the particular case $\sg=0$ and $ \pa V/\pa \phi=0$, are invariant under the  scale-factor duality transformations: 
\beq
a_i \ra a_i^{-1}, ~~~~~~~~ \fb \ra \fb, ~~~~~~~~ \rb \ra \rb, ~~~~~~~~
\pb_i \ra - \pb_i.
\label{18}
\eeq

This type of transformation is called ``dual" as  it generalizes to the case of time-dependent backgrounds the $T$-duality transformation inverting the compactification radius (thus interchanging ``winding" and ``momentum" modes) in the spectrum of a closed string, quantized in the presence of compact spatial dimensions \cite{7}. 
For the invariance under the  transformations (\ref{18}), however, there is no need of a compact geometry; what is required, instead, is a non-trivial transformation of the dilaton. Let us suppose, in fact, that we are inverting a number $n$ of scale factors, say $a_1, \dots, a_n$, with $1 \leq n \leq d$: the condition $\fb \ra \fb$ then implies 
\beq
\phi-  \sum_{i=1}^d \ln a_i = \ti \phi-  \sum_{i=1}^d \ln \ti a_i =
\ti \phi-  \sum_{i=1}^n \ln a_i ^{-1} - \sum_{i= n+1}^d \ln a_i,
\label{19}
\eeq
from which
\beq
\phi \ra \ti \phi= \phi- 2 \sum_{i=1}^n \ln a_i .
\label{20}
\eeq
In the presence of sources, their energy density is also  non-trivially transformed: the condition $\rb \ra \rb$ implies, in fact,
\beq
\r \prod_{i=1}^d a_i= \ti \r \prod_{i=1}^n a_i^{-1} \prod_{i=1+n}^d a_i,
\label{21}
\eeq
from which
\beq
\r \ra 
\ti \r= \r \prod_{i=1}^n a_i^2.
\label{22}
\eeq
The transformation of the pressure is similar, but with an additional ``reflection" of the equation of state along the spatial directions affected by the duality transformation:
\beq
p_i \ra \ti p_i=- p_i \prod_{k=1}^n a_k^2, 
~~~~~~~~~~~~~~~ i= 1, \dots, n. 
\label{23}
\eeq
In any case, given a set of variables $S=\{a_i(t), \phi(t), \r(t), p_i\}$ representing an exact solution of Eqs. (\ref{14})-(\ref{16}), a new  solution can be obtained by inverting an arbitrary number  $n$ (between $1$ and $d$) of scale factors, and is represented by 
\beq
\ti {\cal S}=\{a_1^{-1}, a_2^{-1}, \dots , a_n^{-1}, a_{n+1}, \dots, a_d,~ \ti \phi, ~\ti \r, ~\ti p_1, \dots, \ti p_n,~ p_{n+1}, \dots, p_d\},
\label{24}
\eeq
where $\ti \phi$, $\ti \r$, $\ti p_i$ are given by Eqs. (\ref{20}), (\ref{22}), (\ref{23}), respectively. 

The invariance under the transformations (\ref{18}) is only a particular case of a more general $O(d,d)$ symmetry of the tree-level string cosmology equations \cite{8} (see also the contribution of Meissner \cite{9} to this volume), and can be extended so as to include  the NS-NS two-form $B_{\mu\nu}$ in the effective action. Such an extension is 
also possible  in the presence of fluid sources: a homogeneous gas of strings, in particular, provides a realistic example of source  which is automatically compatible with the $O(d,d)$ symmetry of the background equations \cite{10}. 

In addition, the invariance  under the transformations (\ref{18}) can be extended to the case of non-trivial potentials, $\pa V/\pa \phi \not=0$, and non-zero dilaton couplings to the matter sources, $\sg \not=0$. In both cases, however, we need to generalize those parts of the action describing the self-coupling of the dilaton and the dilaton couplings to the matter fields present in $S_m$. 

In the case of the dilaton potential it is well known \cite{8}-\cite{10} that the invariance under the transformations (\ref{18}) holds for non trivial $V$, provided $V$ depends on $\phi$  through the variable $\fb$. Such a variable, unlike $\phi$, {\em is not} a scalar under general coordinate transformations (as evident from the definition (\ref{13})):  it is thus impossible, in a generic background, to define a potential which is function of $\fb$ and which can be directly inserted as a scalar  into the covariant action (\ref{1}). However, as first pointed out in \cite{11}, the action and the corresponding equations of motion can be written in a generalized form which is invariant under general coordinate transformations in any metric background, using for the potential a non-local  variable which exactly reduces to $\fb$ in the limit of a homogeneous  geometry.  

Here it will be shown that the invariance under the duality transformations (\ref{18}) can be restored also in the presence of the dilaton charge $\sg$, provided the dilaton coupling to the matter sources is parametrized by a non-local variable, as in the case of the potential. This result is new, and will be explicitly derived in the following subsection. 

\subsection{Non-local dilaton interactions}
\label{sec12}

The formalism introduced in \cite{11} is based on the non-local variable $\xi (x)$, defined by 
\beq
\xi(x)\equiv \xi[\phi(x)]=- \ln  \int {\d^{d+1}y\over  \ls^{d}} \left(\sqrt{-g}\, \e^{-\phi} \sqrt{\ep (\na \phi)^2}\right)_y \da(\phi_x-\phi_y),
\label{25}
\eeq
where we have explicitly inserted the parameter
\beq
\ep= {\rm sign}\{ (\na \phi)^2 \}= 
\left\{
\begin{array}{ll}
~~1, ~~~ (\na \phi)^2>0,\\
-1,  ~~~ (\na \phi)^2<0,	
\end{array}  
\right.
\label{26}
\eeq
so as to include in the formalism both time-like and space-like dilaton gradients. Note that we  are using the convenient notation in which  an index appended to round brackets, $( \dots )_x$, means that all quantities inside the brackets are functions of the appended variable. Similarly, $\phi_x \equiv \phi(x)$. We can immediately check that, for a homogeneous background of the type (\ref{8}) with spatial sections of finite comoving volume ($\int \d^d y =V_d=$ const $<\infty$), the variable $\xi$ exactly reduces to the variable $\fb$ of Eq. (\ref{13}). In that case, in fact, an explicit integration gives 
\beq
\xi= \phi- \ln \prod_ia_i -\ln \left(V_d\over \ls^d\right),
\label{27}
\eeq
and the constant volume factor can be simply absorbed by rescaling $\phi$, so that $\xi \equiv \fb$. 

Let us now suppose that the matter couplings and the self-coupling of the dilaton are both parametrized by $\xi$, according to the effective action 
\bea
S=&-&{1\over 2 \ls^{d-1}} \int \d^{d+1}x \,\sqrt{-g}\, \e^{-\phi} \left[R+(\na \phi)^2 +V(\e^{-\xi})\right] 
\nonumber \\ &+&
 \int \d^{d+1}x~ \sqrt{-g}~ {\cal L}_m(\e^{-\xi})+ S_\Sg,
\label{28}
\eea
which is a (generally-covariant) scalar functional of the non-local variable $\xi$. Note that, without loss of generality, we have written both the potential and the matter Lagrangian $\cl$ as a function of $\exp(-\xi)$. In higher-dimensional manifolds with compact spatial sections, in fact, the exponential of the shifted dilaton plays the role of a ``dimensionally-reduced" coupling parameter, and we may thus  expect (at least in a perturbative regime) that dilaton interactions appear as a power expansion (or as a a simple function) of such an exponential \cite{11}. 

The generalized equations of motion  can now be obtained by computing the functional derivative of the action (\ref{28}) with respect to $g^{\mu\nu}$ and $\phi$. The  derivative with respect to the metric, using the standard definition of gravitational stress tensor,
Eq. (\ref{4}), and the properties of the delta distribution, leads to the (integro-differential) equations of motion
\bea
&&
G_{\mu\nu}+\nabla_\mu \nabla_\nu \phi  + {1\over 2} g_{\mu\nu} 
\left(\na \phi^2 -2 \na^2 \phi -V\right)
\nonumber \\ &&
~~~~~~
-{1\over 2} \ga_{\mu\nu} \rp \left(\e^{-\phi} I_V - 2 \ls^{d-1} I_m\right)
=\ls^{d-1}\e^\phi ~T_{\mu\nu},
\label{29}
\eea
which generalize Eq. (\ref{3}) (see Appendix A for the details of the derivation). Here
\bea
&&
\ga_{\mu\nu} = g_{\mu\nu} - {\na_\mu \phi \na_\nu \phi \over (\na \phi)^2},
\label{30}\\
&&
I_V(x)= \ls^{-d} \int \d^{d+1} y \left( \rg \,V'\right)_{y} \da(\phi_{y}-\phi_x),
\label{31}\\
&&
I_m(x)= \ls^{-d} \int \d^{d+1} y \left( \rg \,\cl'\right)_{y} \da(\phi_{y}-\phi_x),
\label{32}
\eea
where the prime denotes the derivative with respect to the argument $\exp(-\xi)$, namely: 
\beq
V'= {\pa V \over \pa(\e^{-\xi})}=-\e^\xi\, {\pa V\over \pa \xi}, 
~~~~~~~~~~~~~~
\cl'= {\pa \cl \over \pa(\e^{-\xi})}=-\e^\xi\, {\pa \cl\over \pa \xi}. 
\label{33}
\eeq
The functional derivative with respect to $\phi$ leads to the dilaton equation of motion, 
\bea
&&
R+ 2 \nabla^2 \phi-(\nabla \phi)^2+V
+\ep {\ga_{\mu\nu} \na^\mu \na ^\nu \phi\over \rp}
\left(\e^{-\phi} I_V - 2 \ls^{d-1} I_m\right)
\nonumber \\ &&
+\left(\e^{-\xi} - \e^{-\phi} J \right) 
\left(V' - 2 \ls^{d-1} \e^\phi \cl'\right)=0,
\label{34}
\eea
generalizing Eq. (\ref{5}). Here 
\beq
J(x)= \ls^{-d} \int \d^{d+1} y \left( \rg \,\rp\right)_y \da'(\phi_x-\phi_y),
\label{35}
\eeq
where $\da'$ denotes the derivative of the delta function with respect to its argument (see Appendix A). The combination of Eqs. (\ref{29}) and (\ref{34}) finally leads to the equation
\bea
&&
R_{\mu\nu}+\nabla_\mu \nabla_\nu\phi 
\nonumber \\ &&
+ {1\over 2}
\left(\e^{-\phi} I_V - 2 \ls^{d-1} I_m\right)
\left(\ep \, g_{\mu\nu}  {\ga_{\a\b} \na^\a \na ^\b \phi\over \rp}
 -\ga_{\mu\nu} \rp\right)
\nonumber \\ &&
+ {1\over 2}g_{\mu\nu}\left(\e^{-\xi} - \e^{-\phi} J \right)
\left(V' - 2 \ls^{d-1} \e^\phi \cl'\right)
 = \ls^{d-1} \e^{\phi}\,T_{\mu\nu}, 
\label{36}
\eea
generalizing Eq. (\ref{7}). 

We can easily check that these new equations, written for a  homogeneous background, are invariant under scale-factor duality transformations  even in the presence 
of non-trivial  potentials and dilaton couplings, i.e. for $\pa V/\pa \xi \not=0$, 
$\pa \cl/\pa \xi \not=0$. Consider, for instance, the  background configuration of Eq. (\ref{8}) with time-like dilaton gradients, for which  $\ep=1$. From Eq. (\ref{30}) we obtain: 
\beq
\ga_0^0=0, ~~~~~~~~~~~~~~~
\ga_i^j = \da_i^j.
\label{37}
\eeq
The $(0,0)$ component of Eq. (\ref{29}) thus coincides with the 
$(0,0)$ component of Eq. (\ref{3}), and is given by Eq. (\ref{9}), as before. 

For the spatial components we first note that, performing the homogeneous limit in which $\xi \ra \fb$, we are lead to the identities 
\bea
\rp  \left(\e^{-\phi} I_V - 2 \ls^{d-1} I_m\right) &\equiv& \e^{-\xi}
\left(V' - 2 \ls^{d-1} \e^\phi \cl'\right) 
\nonumber \\ 
&\longrightarrow& -\left({\pa V\over \pa \fb} - 
 2 \ls^{d-1} \e^\phi{\pa \cl\over \pa \fb}\right);
 \label{38} \\
 {\ga_{\a\b} \na^\a \na ^\b \phi\over \rp}
  \left(\e^{-\phi} I_V - 2 \ls^{d-1} I_m\right) &\equiv& \e^{-\phi}
J \left(V' - 2 \ls^{d-1} \e^\phi \cl'\right) 
\nonumber \\ 
&\longrightarrow& -{\sum_i H_i \over \fpu} 
\left({\pa V\over \pa \fb} - 2 \ls^{d-1} \e^\phi{\pa \cl\over \pa \fb}\right).
\nonumber \\ &&
 \label{39} 
\eea
Using such identities we find that the dependence on $V'$ and $\cl'$ completely disappears from the spatial components of Eq. (\ref{36}) with $\ep=1$, and we obtain the condition
\beq
R_i\,^j+\na_i\na^j \phi= \ls^{d-1} \e^\phi\, T_i\,^j.
\label{40}
\eeq
Written explicitly, the new spatial equation is given by 
 \beq
 \dot H_i -H_i \left(\fpu - \sum_k H_k\right)  =   \ls^{d-1}\, \e^\phi \,  p_i,
 \label{41}
 \eeq
and is thus crucially simplified with respect to the corresponding (local) spatial equation  (\ref{10}). 

The dilaton equation (\ref{34}) also simplifies in the homogeneous limit, thanks to the identities (\ref{38}) and (\ref{39}) from which we obtain
\beq
R+ 2 \nabla^2 \phi-(\nabla \phi)^2+V- {\pa V\over \pa \fb}= -2\ls^{d-1} \e^{\phi} \,{\pa \cl \over \pa \fb}.
\label{42}
\eeq
The explicit form is:
\bea
&&
2 \fpp - \fpu^2 +2 \fpu \sum_iH_i- \sum_i\left(2\dot H_i+H_i^2\right)
- \left( \sum_i H_i\right)^2 +
\nonumber \\ &&
+V(\fb)- {\pa V\over \pa \fb}=\ls^{d-1}\, \e^\phi \, \sg(\fb), 
 \label{43}
 \eea
where we have defined, by analogy with Eq. (\ref{6}), 
\beq 
\sg(\fb)= -2 {\pa \cl \over \pa \fb}.
\label{44}
\eeq

The new set of equations (\ref{9}), (\ref{41}), (\ref{43}) is compatible with scale-factor duality  for any $V$ and $\sg$, as can be shown by rewriting the equations in terms of the shifted variables of Eq. (\ref{13}). With such variables, Eqs. (\ref{9}), (\ref{41}), (\ref{43}) become, respectively,
\bea
&&
\fbp^2-\sum_i H_i^2-V=2 \ls^{d-1}\,\e^{\fb}~ \rb, 
\label{45}\\ &&
\dot H_i- H_i \fbp= \ls^{d-1}\,\e^{\fb}\, \pb_i,
\label{46}\\ &&
2 \fbpp - \fbp^2 -\sum_i H_i^2+V(\fb)- {\pa V\over \pa \fb}=\ls^{d-1}\, \e^{\fb} ~ \sgb (\fb). 
\label{47}
\eea
They are manifestly invariant under the generalized transformations 
\beq
a_i \ra a_i^{-1}, ~~~~~~ \fb \ra \fb, ~~~~~~ \rb \ra \rb, ~~~~~~
\pb_i \ra - \pb_i, ~~~~~~ \sgb \ra \sgb,
\label{48}
\eeq
preserving the shifted version of the dilaton-charge density $\sgb$.

\subsection{The pre-big bang scenario}
\label{sec13}

Let us come back to the cosmological applications of scale-factor duality. Even without using its non-local extensions, the duality symmetry of the equations allows introducing a ``dual complement" of the standard cosmological solutions, and suggests  new possible scenarios for the primordial evolution of our Universe. 

For a simple illustration of this possibility it will be enough to  consider a homogeneous, isotropic and spatially flat metric background, sourced by a barotropic perfect fluid with equation of state $p/\r= \ga=$ const, with negligible dilaton charge. By imposing $\sg=0$, $V=0$, and assuming $\ga \not=0$, one easily finds that Eqs. (\ref{45})-(\ref{47}) are satisfied by the following particular exact solution
\beq
 a=\left(t\over t_0\right)^{2\ga \over 1+d \ga^2}, ~~~~~
 \rb= \r_0 a^{-d \ga}, ~~~~~
 \fb= -{2\over 1+ d \ga^2} \ln \left(t\over t_0\right) +{\rm const},
 \label{49}
 \eeq
 where $t>0$, and $t_0$, $ \r_0$ are positive integration constants. In terms of the non-shifted variables:  
\bea
&&
 a=\left(t\over t_0\right)^{2\ga \over 1+d \ga^2}, ~~~~~~~~~
 \r= \r_0 a^{-d (1+\ga)}, ~~~~~~~~~ p= \ga \r,
 \nonumber \\ &&
 \phi= 2{d \ga -1\over 1+ d \ga^2} \ln \left(t\over t_0\right) +{\rm const},
~~~~~~~~~~~~~~~~~~~~~ t>0.
\label{50}
 \eea
This solution, defined over the real positive semi-axis $t>0$, describes a Universe evolving from a past curvature singularity at $t \ra 0_-$ to an asymptotically flat configuration at $t \ra + \infty$. For $\ga>0$ we have a phase of decelerated expansion and decreasing curvature,
\beq
\dot a >0, ~~~~~~~~ \ddot a <0, ~~~~~~~~ \dot H <0, 
\label{51}
\eeq
typical of the standard cosmological scenario. Also, for a ``realistic" equation of state with $\ga d \leq 1$,  the dilaton turns out to be non-increasing ($\fpu \leq 0$); in particular, for a radiation fluid with $\ga=1/d$, one recovers the radiation-dominated solution at constant dilaton, 
\beq
p= {\r\over d}, ~~~~~~~
 a=\left(t\over t_0\right)^{2 \over 1+d}, ~~~~~~~
 \r= \r_0 a^{-(1+d)}, ~~~~~~~ \phi= {\rm const},
 \label{52}
 \eeq
 which is also an exact solution of the standard Einstein equations. 
 
Thanks to the symmetries of the string cosmology equations we can now obtain  new, different solutions (which have no analogue in the context of the Einstein equations) by performing  a time reflection $t \ra -t$ and, simultaneously, a  dual transformation defined by Eq. (\ref{18}). Starting in particular from Eq. (\ref{50}) we are lead to the background
\bea
&&
 a=\left(-{t\over t_0}\right)^{-{2\ga \over 1+d \ga^2}}, ~~~~~~~~~
 \r= \r_0 a^{-d (1-\ga)}, ~~~~~~~~~ p= -\ga \r,
 \nonumber \\ &&
 \phi= -2{1+d \ga \over 1+ d \ga^2} \ln \left(-{t\over t_0}\right) +{\rm const},
~~~~~~~~~~~~~~~~~~~ t<0, 
\label{53}
\eea
which is still a particular exact solution of  Eqs. (\ref{45})-(\ref{47}). It is  defined on the negative real semi-axis $t<0$, and for $\ga>0$ it  describes a phase of accelerated (i.e. inflationary) expansion and growing curvature: 
\beq
\dot a >0, ~~~~~~~~ \ddot a >0, ~~~~~~~~ \dot H >0.  
\label{54}
\eeq
In this case the Universe evolves from an asimptotically flat initial configuration at $t \ra -\infty$ towards a curvature singularity at $ t \ra 0_-$. The dilaton is always growing  ($\fpu>0$) for $ t \ra 0_-$, even if we consider the dual of the radiation-dominated solution (\ref{52}): 
\bea
&&
p=- {\r\over d}, ~~~~
 a=\left(-{t\over t_0}\right)^{-{2 \over 1+d}}, ~~~~
 \r= \r_0 a^{-d+1}, ~~~~~\phi= -{4d\over d+1}\ln \left(-{t\over t_0}\right).
 \nonumber \\ &&
 \label{55}
 \eea

This interesting property of the low-energy string-cosmology equations -- i.e. the presence of an inflationary ``partner" associated to any standard decelerated solution -- is also valid in the absence of sources. Consider, for instance,  Eqs. (\ref{45})-(\ref{47}) with $p=0=\r$, and $V=0$. In the isotropic limit we find the particular exact solution  
\beq
 a=\left(t\over t_0\right)^{1/\sqrt{d}}, ~~~~~~~~
 \phi= \left(\sqrt d -1\right) \ln \left(t\over t_0\right), 
 ~~~~~~~~ t>0,
 \label{56}
 \eeq
describing decelerated expansion and decreasing curvature. 
By applying the  transformations (\ref{18}) we are lead to the dual solution 
\beq
 a=\left(-{t\over t_0}\right)^{-1/\sqrt{d}}, ~~~~~~
 \phi= -\left(\sqrt d +1\right) \ln \left(-{t\over t_0}\right), 
 ~~~~~~~ t<0, 
 \label{57}
 \eeq 
 describing accelerated expansion, growing curvature and growing dilaton. Actually, both the vacuum and the fluid-dominated solutions can be obtained as asymptotic limits of the general exact solution of the system of equations (\ref{45})-(\ref{47}) (for barotropic sources, with $V=0$ and $\sg=0$), in the large-curvature and small-curvature limits, respectively \cite{12,12a}. 

It is well known that the decelerated configurations, typical of the standard cosmological scenario, cannot be extended back in time without limits: the range of the time coordinate is bounded from below by the presence of the initial singularity (indeed, going back in time, the growth of $H$ is unbounded, and the curvature blows up to infinity in a finite proper-time interval).

The standard scenario, however, is certainly incomplete because it excludes inflation. The inclusion of inflation, on the other hand, modifies the behavior of the curvature scale: during a phase of ``slow-roll" inflation \cite{13}, for instance, the background geometry can be approximately described by a de Sitter-like metric where $\dot H \simeq 0$, and in which the curvature tends to settle at a constant. One might think, therefore, that a complete (and realistic)  cosmological scenario could avoid the initial singularity, replacing it with a primordial inflationary phase at constant curvature. 

Unfortunately, however, an epoch of accelerated expansion at constant curvature, described by the Einstein equations, and dominated by the potential energy of some ``inflaton" scalar field satisfying causality and weak-energy conditions, cannot be ``past eternal", as proved in \cite{14}. Thus, the conventional inflationary scenario mitigates the rapid growth of the curvature typical of the standard cosmological evolution, and shifts back in time the position of the initial singularity, without completely removing it, however (namely, without extending in a geodesically complete way the model, back in time, to infinity). 

If a constant-curvature  phase is not appropriate to construct a regular  model (fully extended over the whole temporal axis), the alternative we are left is a model in which the curvature, as we go  back in time, after reaching a maximum, at some point,  starts decreasing; in other words, a model in which the standard evolution is completed and complemented by a primordial phase with a specular behavior of $H$ with respect to the standard one. Remarkably, this is exactly what can be obtained assuming that the cosmological evolution satisfies a principle of ``self-duality" -- i.e. assuming that the past evolution of our Universe is described by the ``dual complement" of the  present one \cite{2,19}. 

More precisely, if we consider a cosmological model satisfying (at least approximately) the self-dual condition $a(t)= a^{-1}(-t)$, such that the standard decelerated regime at $t>0$ smoothly evolves, back in time, into the accelerated partner at $t<0$, we can then obtain a scenario in which the singularity is automatically regularized, and the initial evolution is automatically of the inflationary type. In such a context the big bang singularity is replaced by an epoch of high (but finite) curvature, characterizing the transition between the standard cosmological phase ($\dot H <0$) and its dual ($\dot H >0$): it comes natural, in such a context, to call {\em ``pre-big bang"} the initial  phase ($t<0$) at growing curvature and growing dilaton, in contrast to the subsequent {\em ``post-big bang"} phase ($t>0$),  describing the standard cosmological evolution. 

\begin{figure}
\centering
\includegraphics[height=9cm]{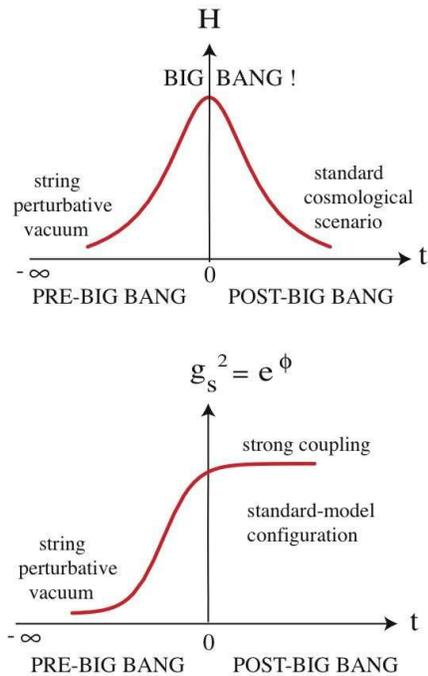}
\caption{Qualitative time-evolution of the curvature scale (upper panel) and of the string coupling (lower panel), for a typical self-dual background which smoothly interpolates between the pre-big bang and the post-big bang phase, starting from the string perturbative vacuum.}
\label{f2}      
\end{figure}

The dilaton, on the other hand,  provides an exponential parametrization of the (tree-level) string coupling $g_s= \exp(\phi/2)$, controlling the relative strength of all (gravitational and gauge) interactions \cite{3,4}. The principle of self-duality thus suggests that the Universe is lead to its present state after a long evolution started from an extremely simple  -- almost trivial -- configuration, characterized by a nearly flat geometry and by a very small coupling parameter, 
\beq
H^2 \ra 0, ~~~~~~~~~~~~
 \phi \ra -\infty,  ~~~~~~~~~~~~
g_s^2 = \e^\phi \ra 0,
\label{58}
\eeq
the so-called string perturbative vacuum (see Fig. \ref{f2}). In this case, the initial Universe is characterized by a regime of extremely low energies in which the ``curvatures" (i.e. the field gradients) are small ($\ls^2 H^2 \ll1$, $\ls^2 \fpu^2 \ll1$, \dots), the couplings are weak ($g_s^2 \ll1$), and the background dynamics can be appropriately described by the lowest-order string effective action, at  tree-level in the $\ap$ and quantum loop expansion (also in agreement with the hypothesis of ``asymptotic past triviality" \cite{15}). We can talk of ``birth of the Universe from the string perturbative vacuum", as also  pointed out in a quantum cosmology context (see e.g \cite{GRF,20}. 

This picture is in remarkable contrast with the standard (even inflationary) picture in which the Universe starts evolving from a 
highly-curved geometric state: the more we go back in time, in that context, the more we enter a Planckian and (possibly) trans-Planckian \cite{16} non-perturbative regime of ultra-high energies, requiring the full inclusion of quantum gravity effects, to all orders, for a correct description. 

The principle of self-duality, on the contrary, suggests a picture in which the more we go back in time (after crossing the epoch of maximal curvature), the more we approach a {\em flat, cold} and {\em vacuum} configuration (strongly reminiscent of the ``biblical" scenario quoted at the beginning of Sect. \ref{sec1}), which can be appropriately described by the classical background equations obtained from the action (\ref{1}). Quantum effects, in the form of higher-curvature and higher-loop contributions, are expected to become important only {\em towards the end} of the pre-big bang phase, when the background approaches the string scale at $t \ra 0_-$. Actually, all studies performed so far have  shown that such corrections {\em must} become dominant, eventually, in order to stop the growth of the curvature \cite{17} and possibly trigger a smooth transition to the post-big bang regime \cite{18}. 

\subsection{A smooth ``bounce"}
\label{sec14}

The lowest-order string effective action can appropriately describe the phase of primordial background evolution typical of the  pre-big bang scenario, but not the transition to the standard decelerated regime  occurring at high curvatures and strong coupling, and requiring the introduction of higher-order corrections. Referring the reader to the existing literature for a detailed review of the transition models studied so far (see for instance \cite{19}),  we shall  present here only two simple phenomenological examples, by applying, to this purpose, the formalism introduced in subsection \ref{sec12} (and Appendix A). In these examples, in fact, the bouncing transition is induced by 
 the presence of a non-local effective potential $V(\fb)$, expected to simulate the backreaction of the quantum loop corrections in higher-dimensional manifolds with compact spatial sections \cite{11}. 
 
\begin{figure}
\centering
\includegraphics[height=5cm]{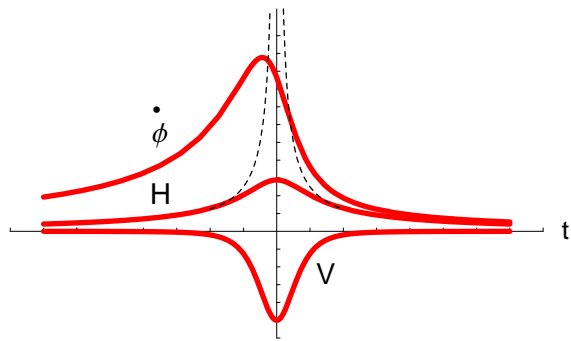}
\caption{Plot of the curvature, of the dilaton kinetic energy, and of the potential $V(\fb)$, for the bouncing solution (\ref{60}). The dashed curves represent the (singular) vacuum solutions (\ref{56}), (\ref{57}), obtained with $V=0$. All curves are plotted for $t_0=1$, $V_0=1$, and $d=3$.}
\label{f3}      
\end{figure}

The first example is based on a potential which, in the homogeneous limit, takes the form
\beq
V(\fb)= -V_0 e^{4\fb}, ~~~~~~~~~~~~~~~~~ V_0>0,
\label{59}
\eeq
and which may thus perturbatively interpreted as a four-loop potential.  With this potential, the duality-invariant equations (\ref{45})--(\ref{47}), in vacuum ($\r=p=\sg=0$), and in the isotropic limit, are solved by the particular exact solution \cite{20}:
\bea
&&
a(t)= a_0 \left[{t\over t_0} +\left(1+{t^2\over t_0^2}\right)^{1/2} \right]^{1/\sqrt d},
\nonumber \\ &&
\fb= -{1\over 2} \ln \left[t_0 \sqrt{V_0} \left(1+{t^2\over t_0^2}\right)\right]+ {\rm const},
\nonumber \\ &&
\phi= \ln {\left[t/t_0+\left(1+{t^2/ t_0^2}\right)^{1/2}\right]^{\sqrt{d}}
\over \left(1+{t^2/ t_0^2}\right)^{1/2}} +{\rm const},
\label{60}
\eea
where $t_0$ and $a_0$ are positive integration constants. This regular  ``bouncing" solution is exactly self-dual -- as it satisfies $a(t)/a_0= a_0/a(-t)$ -- and is characterized by a bounded, ``bell-like" shape of the curvature and of the dilaton kinetic energy (see Fig. \ref{f3}). The solution smoothly interpolates between the pre-and post-big bang vacuum solutions (\ref{57}), (\ref{56}) (corresponding to the dashed curves of Fig. \ref{f3}), which are recovered in the asymptotic limits $t \ra -\infty$ and $t \ra + \infty$, respectively. The bounce of the curvature, and the smooth transition between the two branches of the low-energy solutions, is induced and controlled by the potential (\ref{59}) which dominates the background evolution in the high-curvature limit $|t| \ra 0$, and which becomes rapidly negligible as $t \ra \pm \infty$, as illustrated in Fig. \ref{f3}. 

It should be noted that in this solution the dilaton keeps growing, monotonically, even in the limit $t \ra +\infty$. In more realistic examples, however, such a growth is expected to be damped by the interaction with the matter/radiation post-big bang sources \cite{20a}, and/or by the action of a suitable non-perturbative potential appearing in the strong coupling regime. 

The second example of bounce is based on a general integration of the duality-invariant equations (\ref{45})--(\ref{47}), in the presence of isotropic fluid sources with $\sg=0$ and of a two-loop (non-local) potential which in the homogeneous limit takes the form
\beq
V(\fb)= -V_0 e^{2\fb}, ~~~~~~~~~~~~~~~~~ V_0>0. 
\label{61}
\eeq
In this case the equations can be integrated exactly not only for barotropic equations of state ($p/\r=\ga=$ const), but also for any ratio $p/\r$ which is an integrable function of an appropriately defined time-like parameter \cite{2}. 

\begin{figure}
\centering
\includegraphics[height=5cm]{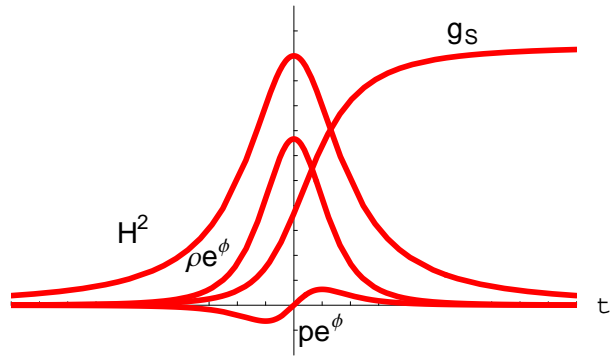}
\caption{Plot of the curvature, of the string coupling, of the effective energy density and of the effective pressure for the self-dual solution (\ref{64}). The curves are plotted for $d=3$, $L=1$, $x_1=1$, $\phi_0=0$, and $a_0=\exp(-2/3)$.}
\label{f4}      
\end{figure}

An interesting example (motivated by the study of the equation of state of a string gas in rolling backgrounds \cite{21}) is the case in which $p/\r$ smoothly evolves from the value $\ga=-1/d$ at $t=-\infty$ to the value $\ga=1/d$ at $t=+\infty$, thus connecting the radiation equation of state to its dual partner, according to the law:
\beq
{p\over \r}={1\over d} {x\over \sqrt{x_1^2+x^2}}.
\label{62}
\eeq
Here $x_1$ is an arbitrary integration constant, and $x$ is a (dimensionless) time-like coordinate defined by 
\beq
{\d x \over \d t}= {L\over 2} \rb,
\label{63}
\eeq
where $L$ is a constant with dimensions of length (we are using units in which $2 \ls^{d-1}=1$, so that $[\r]= L^{-2}$). Using Eqs. (\ref{61})--(\ref{63}), and choosing a simplifying set of integration constants (appropriate to the pedagogical purpose of this paper), we can then obtain the following particular exact solution \cite{2},
\bea
&&
a=a_0 \left(x+ \sqrt{x^2+x_1^2}\right)^{2/ (d-1)},
\nonumber \\ &&
\e^\phi=a_0^d \e^{\phi_0}\left(1+{x\over \sqrt{x^2+x_1^2}}\right)^{2d/(d-1)}, 
\nonumber\\ &&
\r \e^\phi= {d-1\over dL^2} \e^{2\phi_0} \left(x^2+x_1^2\right)^{-{(d+1)/(d-1)}}, 
\nonumber\\ &&
p\e^\phi= {d-1\over d^2L^2} \e^{2\phi_0}x \left(x^2+x_1^2\right)^{-{(3d+1)/2( d-1)}}, 
\label{64}
\eea
where $a_0$ and $\phi_0$ are integration constants. The smooth and bouncing behavior of this solution is illustrated in Fig. \ref{f4}. 

The above solution is self-dual, in the sense that $\fb(x)=\fb(-x)$, $\rb(x)=\rb(-x)$, and
\beq
\left[a(x)\over a_0 x_1^{2/(d-1)}\right]=
\left[a(-x)\over a_0 x_1^{2/(d-1)}\right]^{-1}
 \label{65}
 \eeq
(with an appropriate choice of the integration constant $a_0$ it is always possible to set to $1$ the fixed point of scale-factor inversion). The solution satisfies, asymptotically, 
\bea
&&
x \ra - \infty ~~~~~\Rightarrow ~~~~~~~ a \sim (-x)^{-2/(d-1)} \sim \rb \sim {\d x\over \d t}, 
\nonumber\\ &&
x \ra +\infty ~~~~~\Rightarrow ~~~~~~~ a \sim x^{2/(d-1)}~~~~~~ \sim {1\over \rb} \sim {\d t\over \d x}.  
\label{66}
\eea 
Re-expressing $a$, $\phi$, $\r$, $p$, in the asymptotic limits $x \ra \pm \infty$ in terms of the cosmic time $t$, we can check that this solution smoothly interpolates between the pre-big bang configuration (\ref{55}) describing accelerated expansion, growing dilaton, negative pressure, and the final post-big bang configuration (\ref{52}), describing the radiation-dominated state with frozen dilaton and decelerated expansion. As in the previous case the smoothing out of the tree-level singularity, and the appearance of bouncing transition, is a consequence of the effective potential (\ref{61}). 

\subsection{Cosmological perturbations}
\label{sec15}

The phase of pre-big bang evolution, being accelerated, can  amplify the quantum fluctuations of the metric tensor (and of other background fields) just like any other type of inflationary evolution. However, because of the kinematic properties of pre-big bang inflation (associated to the shrinking of the Hubble horizon $H^{-1}$), 
the spectral distribution of the  metric fluctuations, after their amplification, tends to grow with frequency \cite{22}. This peculiar aspect of the spectrum may be regarded as representing both an advantage and a difficulty of pre-big bang models with respect to other models of inflation. 

The advantage is of phenomenological nature, and refers to the  transverse and traceless tensor part of the metric fluctuations. Their amplification leads to the formation of a stochastic background of relic gravitational waves whose spectral energy density, $\Om_g$, grows wih frequency:
\beq
\Om_g(\om,t)= \left(H_1\over \Mp\right)^2 \Om_r(t) \left( \om \over \om_1\right)^\da,  ~~~~~~~~~~~
\da>0, ~~~~~~~~ \om \leq \om_1.
\label{69}
\eeq
Here $\Mp=8\pi G= \lp^{-1}$ is the Planck mass, $H_1\simeq \Ms=\ls^{-1}$ the inflation-radiation transition scale (expected to be controlled by the string mass scale $\Ms $), $\Om_r= \r_r/\r_c$ is the fraction of critical energy density in radiation, $\om_1$ is the ultraviolet cut-off (i.e. the maximal amplified frequency) of the spectrum, and $\da$ a model-dependent parameter depending on the background kinematics \cite{22,23,24} (see also the contribution of Buonanno and Ungarelli \cite{25} to this volume). 

Thanks to the growth of the spectrum, the cosmic graviton background present today as a relic of the inflationary epoch is higher at higher frequencies (in particular, higher than the backgrounds predicted by conventional models of inflation), and thus more easily detectable by current gravitational antennas (see e.g. \cite{19}). Conversely, however, the spectrum is strongly suppressed in the low-frequency regime: we should thus expect, in particular, a negligible contribution of tensor metric perturbations to the observed CMB anisotropy on large scales (as in the case of the ekpyrotic \cite{ek} and ``new ekpyrotic \cite{nek} scenarios where, however, the gravitational background is expected to be low even in the low-frequency regime \cite{grav}). 
It may be stressed, in this connection, that the possible absence of tensor contributions at large scales emerging from (planned) future measurements of the CMB polarization (such as those of WMAP, PLANCK), in combination with a positive signal possibly detected at high frequency by the next generation of gravitational antennas (such as LIGO/VIRGO, LISA, BBO, DECIGO), could represent a strong experimental signal in favor of models of pre-big bang inflation (see e.g. \cite{26}). 

The difficulties associated to a growing spectrum refer  to the scalar part of the metric perturbations. In fact, a growing scalar spectrum cannot account for the observed peak structure of the temperature anisotropies of the CMB radiation, which requires, instead, a nearly flat (or ``scale-invariant") primordial distribution:  
$\Om_s(\om) \sim \om^{n_s-1}$, with $n_s \approx 1$. There are two possible ways out of this problem. 

A first possibility relies on the growth of the dilaton -- and thus of the string coupling $g_s^2= \exp \phi$ -- during the phase of pre-big bang inflation. Even starting at weak coupling, a pre-big bang background  unavoidably evolves towards the strong coupling regime $g_s  \sim 1$. If the bounce is not immediate then the Universe, before the transition to the  standard regime, enters a strong-coupling phase where higher-dimensional extended objects like Dirichlet branes and antibranes \cite{4} (whose tension is proportional to the inverse of the string coupling) become light, and can be copiously produced \cite{27}. The cosmic evolution may become dominated by the presence of these higher-dimensional sources \cite{28} and, in that context, a phase of conventional slow-roll inflation can be triggered by the interaction (and eventual collision) of a brane-antibrane pair \cite{29}  (see also the contribution of Tye \cite{30} to this volume). This new inflationary regime may efficiently dilute all pre-existing inhomogeneities and generate a new spectrum of scale-invariant, adiabatic scalar perturbations, as required for a successful explanation of the observed anisotropy. This may resolve the incompatibility between a (growing)  spectrum of pre-big bang perturbations and present large-scale observations.  

There is, however, a second possibility which avoids introducing additional inflationary epochs besides the initial dilaton-dominated one, and which is based on the so-called ``curvaton mechanism" \cite{31}. According to this mechanism the (flat, adiabatic) spectrum of scalar metric perturbations, responsible for the observed anisotropies,  is not produced during the primordial evolution: instead, it is the outcome of the post-inflationary decay of a massive scalar field (the curvaton), whose quantum fluctuations are amplified during inflation with a nearly flat spectrum, and are converted into curvature perturbations after its decay.  In the context of the pre-big bang scenario the role of the curvaton is possibly played by the Kalb-Ramond axion $\sg$ \cite{32}, associated -- by space-time duality -- to the four components of the NS-NS two-form $B_{\mu\nu}$ present in the massless multiplet of the string spectrum. 

For a brief discussion of this possibility we should explain, first if all, why axion fluctuations can be amplified by pre-big bang inflation with a flat spectrum \cite{33}, unlike metric fluctuations. The reason is that the slope of the spectrum is directly related to kinematic behavior of the effective ``pump field" responsible for the amplification, and that metric and axion fluctuations have different pump fields, even in the same given background. 

In order to clarify this point let us complete the low-energy action (\ref{1}) by adding the contribution of the antisymmetric field $B_{\mu\nu}$, considering (for simplicity) a model already dimensionally reduced to four space-time dimensions:
\bea
&&
S=-{1\over 2\ls^{2}} \int_{\Om} \d^{4} x \sqrt{-g} \, \e^{-\phi} \left(R+ \pa_\mu\phi\,\pa^\mu \phi +V-{1\over 12} H_{\mu\nu \a} H^{\mu\nu\a}\right),
\nonumber\\ &&
~~~~~
H_{\mu\nu\a}= \pa_\mu B_{\nu\a}+ \pa_\nu B_{\a\mu}+ \pa_\a B_{\mu\nu}.
\label{70}
\eea
In the absence of sources the equations of motion for $B_{\mu\nu}$ are automatically satisfied by introducing the ``dual" axion field $\sg$, such that
\beq
H^{\mu\nu\a}= {\e^\phi\over  \sqrt{-g}}\, \ep^{\mu\nu\a\b} \pa_\b \sg,
\label{71}
\eeq
and the last term of the action (\ref{70}) can be replaced by
\beq
S={1\over 4 \ls^2}\int \d^4x \sqrt{-g}~\e^{\phi}(\na \sg)^2. 
\label{72}
\eeq
Perturbing the metric and the axion field,
\beq
g_{\mu\nu} \ra g_{\mu\nu}+ h_{\mu\nu}, ~~~~~~~~~~
\sg \ra \sg + \da \sg,
\label{73}
\eeq
around a homogeneous, conformally flat metric background, using the conformal time coordinate $\eta$ (such that $\d t=a\d \eta$), and applying the standard formalism of linear cosmological perturbations (see e.g. \cite{34}), we obtain for tensor metric and axion fluctuations, respectively, the following quadratic actions:
\bea
&&
S_h={1\over 2} \int \d^3x\, \d \eta \,z^2_h(\eta) \left(h^{\pr 2} + h \nabla^2 h \right),
\nonumber \\ &&
z_h= {a\over \sqrt 2 \,\ls}\, \e^{-\phi/2},
\label{74}\\
&&
S_\sg={1\over 2} \int \d^3x\, \d \eta \,z^2_\sg(\eta) \left(\da \sg^{\pr 2} + \da \sg \nabla^2 \da \sg \right),
\nonumber \\ &&
z_\sg= {a\over \sqrt 2 \,\ls}\, \e^{\phi/2}.
\label{75}
\eea
Here $h$ is one of the two physical polarization states of tensor perturbations, the primes denote differentiation with respect to $\eta$, and $\nabla^2$ is the flat-space Laplace operator, $\na^2= \da^{ij} \pa_i\pa_j$. The variation of these actions with respect to $h$ and $\da \sg$ leads to the equations of motion, which can be written in terms of the canonical variables $u= (h z_h)$ and $v=(\ds z_\sg)$ as follows:
\bea
&&
(h z_h)''-\left(\na^2+{z_h''\over z_h}\right) (h z_h)=0, 
\label{76}\\ &&
(\ds z_\sg)''-\left(\na^2+{z_\sg''\over z_\sg}\right) (\ds z_\sg)=0. 
\label{77}
\eea
The canonical equations are the same for $u$ and $v$, but the pump fields, $z_h$ and $z_\sg$, are different. 

Consider, for instance, the axion equation (\ref{77}), and recall that during inflation the accelerated evolution of the pump field can be parametrized as a power-law evolution in the negative range of the conformal-time parameter \cite{19, 34}, i.e. 
\beq
z_\sg(\eta)= {\Mp\over \sqrt{2}} \left(-{\eta\over \eta_1}\right)^{\a_\sg}, 
~~~~~~~~~~~~~ -\infty \leq \eta <0,
\label{78}
\eeq
where $\eta_1>0$ is some appropriate reference time-scale. Expanding in Fourier modes, Eq. (\ref{77}) becomes a Bessel equation  for the mode $v_k$, 
\beq
v_k''+\left[k^2 -{(\nu_\sg^2 -1/4)\over \eta^2} \right] v_k=0, 
~~~~~~~~ \nu_\sg={1\over 2}-\a_\sg,
\label{79}
\eeq
and its general solution can be conveniently written as a combination of first-kind and second-kind Hankel functions \cite{35}, of argument $k\eta$ and index $\nu_\sg$, as follows:
\beq
v_k= (-\eta)^{1/2} \left[A_+(k)H_{\nu_\sg}^{(2)}(k\eta)+ 
A_-(k)H_{\nu_\sg}^{(1)}(k\eta)\right].
\label{80}
\eeq

We shall now canonically normalize this general solution by imposing that the initial state of the fluctuations corresponds to a spectrum of quantum vacuum fluctuations \cite{19,34}. More explicitly, we shall require that the mode $v_k$, on the initial spatial hypersurface at $\eta \ra -\infty$, may represent freely oscillating, positive frequency modes satisfying the canonical normalization 
\beq
v_kv_k^{\prime \ast}- v_k'v_k^{\ast}=i,
\label{81}
\eeq
from which
\beq
v_k \ra  {\e^{-ik\eta}\over \sqrt{2k}}, ~~~~~~~~~~~~~
\eta \ra -\infty
\label{82}
\eeq
(modulo an arbitrary phase). Using the large argument limit of the Hankel functions \cite{35},
\beq
H_{\nu}^{(2)}(k\eta)= \sqrt{2\over \pi k\eta}\, \e^{-ik\eta-i\ep_\nu}, ~~~~~~~~~
H_{\nu}^{(1)}(k\eta)= \sqrt{2\over \pi k\eta}\, \e^{ik\eta+i\ep_\nu}
\label{83}
\eeq
($\ep_\nu=-\pi/4-\nu \pi/2$), we obtain $A_+=\sqrt{\pi/4}$ and $A_-=0$.  The normalized exact solution for the the axion fluctuations $\ds_k$ can be finally written as
\beq
\ds_k={v_k\over z_\sg}={\e^{i\theta_k}\over \Mp} 
\left({\pi\eta_1  \over 2}\right)^{1/2} 
\left({\eta  \over \eta_1}\right)^{\nu_\sg} H_{\nu_\sg}^{(2)}(k\eta), 
\label{84}
\eeq
where $\theta_k$ is an arbitrary phase determined by the choice of the initial conditions. 

In order to determine the spectrum of the fluctuations after their inflationary amplification we must then consider the limit $\eta \ra 0_-$, in which $|k \eta| \ll1$ and the amplitude of the mode $k$ is stretched ``outside the horizon". We can use, to this purpose, the small argument limit of the Hankel functions \cite{35}, which reads (for $\nu \not=0$), 
\beq
H_{\nu}^{(2)}(k\eta)= p^\ast_\nu(k\eta)^\nu-iq_\nu(k\eta)^{-\nu}+ \dots
\label{85}
\eeq
where $q_\nu$ and $p_\nu$ are complex ($\nu$-dependent) coefficients (for $\nu=0$ there are additional logarithmic corrections). We obtain, in this limit, 
\beq
\ds_k={v_k\over z_\sg} \ra {\e^{i\theta_k}\over \Mp} 
\left({\pi\eta_1  \over 2}\right)^{1/2} 
\left[-iq_{\nu_\sg} (k\eta_1)^{-\nu_\sg}+p_{\nu_\sg}^\ast (k\eta_1)^{\nu_\sg}
\left(\eta\over \eta_1\right)^{2\nu_\sg}  \right]. 
\label{86}
\eeq
The cases we are interested here are limited to ``conventional" inflationary backgrounds with $\a_\sg \leq 1/2$, i.e. $ \nu _\sg \geq 0$ (see \cite{26} for a detailed discussion of all possibilities). For such backgrounds the time-dependence of $\ds_k$ tends to disappear as $\eta \ra 0_-$, the fluctuations become frozen, asymptotically, and their (dimensionless) spectral amplitude $k^3|\ds_k|^2$, controlling the typical amplitude of the perturbations on a comoving length scale $r= k^{-1}$ \cite{34}, has the following $k$-dependence: 
\beq
 k^3 \left|\ds_k\right|^2 \sim k^{3-2\nu_\sg}= k^{2+2\a_\sg}.
\label{87} 
\eeq
This result also holds in the limiting case $\a_\sg=1/2$ with the only addition of a mild logarithmic correction \cite{23,24}, i.e. 
$ k^3 \left|\ds_k\right|^2 \sim k^3 \ln^2 (k\eta_1)$. 

The above calculations can be exactly repeated, in the same form, for the tensor perturbation variable, starting from Eq. (\ref{76}): the resulting spectrum is formally the same,
\beq
 k^3 \left|h_k\right|^2 \sim k^{3-2\nu_h}= k^{2+2\a_h}, 
\label{88} 
\eeq
with the difference that the spectral slope is now determined by the power $\a_h$, controlling the evolution of the tensor pump field $z_h$ through an equation analogous to Eq. (\ref{78}).

We are now in the position of discussing the possible pre-big bang production of  a flat spectrum of axion fluctuations, even if the associated metric fluctuations are amplified (in the same background) with a growing spectrum. Let us consider, to this purpose, an exact anisotropic solution of the string cosmology equations (\ref{9})--(\ref{11}), in vacuum, and without dilaton potential. The solution describes a phase of pre-big bang inflation  characterized by the accelerated (isotropic) expansion of three spatial dimensions, with scale factor $a(\eta)$, and by the accelerated contraction of $n$ ``internal" spatial dimensions, with scale factors $b_i(\eta)$, $i=1, \dots , n$. In conformal time, such a solution can be parametrized for $\eta \ra 0_-$ as \cite{12a,19} 
\bea
&&
a=\left(-{\eta\over \eta_1}\right)^{\b_0/ (1-\b_0)} , 
~~~~~~~~~~~~~~~~~~~~~~
b_i= \left(-{\eta\over \eta_1}\right)^{\b_i/(1-\b_0)} , 
\nonumber \\ &&
 \phi_{4+n}= {\sum_i \b_i +3 \b_0-1\over 1-\b_0}~\ln 
\left(-{\eta\over \eta_1}\right),
\label{89}
\eea
where the constant coefficients $\b_0$, $\b_i$ satisfy the Kasner-like condition
\beq
\sum_i \b_i^2 +3 \b_0^2=1,
\label{90}
\eeq
and $\phi_{4+n}$ is the higher-dimensional dilaton appearing in the full $(4+n)$-dimensional effective action. The four-dimensional dilaton $\phi$ is related to $\phi_{4+n}$ by
\beq
\e^{-\phi}= V_n \,\e^{-\phi_{4+n}}\equiv \e^{-\phi_{4+n}} \prod_i b_i,
\label{91}
\eeq
namely by
\beq
\phi= \phi_{4+n}-\sum_i \ln b_i= {3 \b_0-1\over 1-\b_0} \ln \left(-{\eta\over \eta_1}\right).
\label{92}
\eeq

Let us compute, for this background, the kinematic powers $\a_h$ and $\a_\sg$ controlling the evolution of the pump fields (\ref{74}), (\ref{75}):
\bea
&&
z_h \sim a \e^{-\phi/2} \sim (-\eta)^{\a_h}, ~~~~~~~~~~~~~~~~
\a_h ={1\over2},
\label{93} \\ &&
z_\sg \sim a \e^{\phi/2} \sim (-\eta)^{\a_\sg}, ~~~~~~~~~~~~~~~~~~
\a_\sg ={5\b_0-1\over2(1-\b_0)}.
\label{94}
\eea
It follows, according to Eq. (\ref{88}), that the spectrum of tensor (as well as of scalar) metric perturbations is always characterized  by a  slope which is cubic (modulo log corrections) \cite{12a,23,24}, and which is also ``universal", in the sense that it is insensitive to the background parameters $(n, \b_0, \b_i)$. For the axion fluctuations, on the contrary, we find from Eq. (\ref{87}) that the spectral slope is strongly dependent on such parameters, and that a scale-invariant spectrum with $2+2\a_\sg=0$ is allowed, in particular, provided 
$\b_0=-1/3$.

We may note, in the special case in which the background is fully isotropic and expanding (i.e., $\b_0=\b_i<0$), that the Kasner condition (\ref{90}) implies $\b_0=-1/\sqrt{d}$, so that a scale-invariant spectrum corresponds to $d=9$, i.e. just to the number of spatial dimensions determined by critical superstring theory \cite{3,4}. 

In the less special case in which the spatial geometry can be factorized as the product of a $3$-dimensional and a $n$-dimensional isotropic subspaces we have, instead, $\b_i=\b \not=\b_0$, and $3 \b_0^2+ n \b^2=1$. The spectral slope, in this case, can be expressed in terms of the parameter
\beq
r={1\over 2} \left(\dot V_n\over V_n\right)\left(\dot V_3\over V_3\right)^{-1}={n\b\over 6\b_0},
\label{93a}
\eeq
controlling  the relative time-evolution of the proper volumes of the  internal and external spaces. Eliminating $\b$ in terms of $\b_0$ through the Kasner condition, and replacing $\b_0$ with $r$ in Eq. (\ref{94}), one can then parametrize the deviations from a flat axion spectrum as the relative shrinking or expansion of the two subspaces \cite{36}. 

Given a sufficiently flat spectrum of axion fluctuations, amplified by the phase of pre-big bang inflation, we are then lead to a post-big bang configuration which is initially characterized (at some given time scale $\eta_i$) by a primordial  sea of ``isocurvature" scalar perturbations, dominated on super-horizon scales by the axion fluctuations $\ds$ (the metric fluctuations are subdominant on such large scales, being strongly suppressed by the steep slope of their spectrum). The axion can play the role of the curvaton provided that  the initial configuration, besides containing the initial fluctuations $\ds_i$,  also contains a non-vanishing axion background, $\sg_i\not=0$, whose energy density $\r_\sg$  -- even if subdominant -- is initially determined by an appropriate potential (possibly approximated by $V_\sg \sim m^2 \sg^2$). In that case the background evolution, after an initial slow-roll regime, leads to a phase where the axion background starts oscillating with proper frequency $m$, at a curvature scale $H\sim m$, simulating a dust fluid ($\r_\sg \sim a^{-3}$) which may become dominant with respect to the radiation fluid, and eventually  decay at the typical scale $H \sim \lp^2 m^3$. 

In such a type of background the axions fluctuations $\ds$ become linearly coupled to scalar metric perturbations, and may act as sources for the so-called Bardeen potential $\Psi$.  New metric perturbations can then be generated, starting from $\Psi(\eta_i)=0$, with the same spectral slope as the axion one, and with a spectral amplitude not smaller, in general, than the axion amplitude. Referring to the literature for a detailed computation \cite{31,32}, we shall recall here that the final spectrum (after the axion decay) of the super-horizon Bardeen potential is related to the initial axion perturbations by 
\bea
&& ~
\left|\Psi_k\right|=\lp f(\sg_i) \left|\ds_k(\eta_i)\right| , 
\nonumber \\ && 
f(\sg_i)=c_1{\Om_\sg\over \lp\sg_i}+ c_2+c_3\, \lp\sg_i
\label{94a}
\eea
(the $\lp$ factors are due to the canonical normalization of the axion field and of its fluctuations). Here $\sg_i$ is the initial amplitude of the axion background, $\Om_\sg \sim 1$ is the axion fraction of critical density at the axion decay epoch, and $c_1$, $c_2$, $c_3$ are dimensionless numbers of order one ($\Om_\sg$ cannot be much smaller than one, to avoid a too strong ``non-Gaussianity" of the spectrum" \cite{39}). Thanks to its structure, the ``form factor" $f(\sg_i)$ has a minimum of order one around $\lp \sg_i \sim 1$. A (nearly) scale-invariant axion spectrum thus reproduces a (nearly) scale-invariant spectrum of scalar metric  perturbations. 

As discussed in the literature, a curvaton-induced spectrum of scalar metric perturbations provides the right ``adiabatic" initial conditions for reproducing the observed temperature anisotropies of the CMB radiation, exactly as in the case of the slow-roll scenario. The only difference is the ``indirect" (i.e., post-inflationary) production of the scalar spectrum, triggered by the presence of a non-vanishing axion background. It must be stressed, however, that the direct connection (\ref{94a}) with the axion spectrum of primordial origin gives us the possibility of extracting, from present CMB observations, important constraints on the parameters of pre-big bang models of inflation \cite{32}. 

In particular, using the experimental normalization of the anisotropy spectrum, and the direct relation between the pre-big bang inflation scale $H_1$ and the string scale $\Ms$, one can speculate about the possibility of ``weighing the string mass with the CMB data" \cite{37}. Another application concerns the slope of the scalar perturbation spectrum which, according to most recent WMAP results \cite{38}, is given by
\beq
n_s \equiv 3+ 2\a= 0.951^{+0.015}_{-0.019}.
\label{95}
\eeq
Using Eq. (\ref{94}), and the Kasner condition (\ref{90}), one obtains 
\beq
\b_0 \simeq -0.355, ~~~~~~~~~~~~~
\sum_i \b_i^2 \simeq 0.62.
\label{96}
\eeq
With $d=9$ dynamical dimensions this result seems to point out the existence of a small anisotropy between the kinematics of the external and internal spaces during pre-big bang inflation (a fully isotropic expansion would correspond, in fact, to $\b_0=-1/\sqrt 9 \simeq -0.33$ and $\sum_i \b_i^2 =6/9 \simeq 0.66$). It should be noted, however, that other interpretations of the data are also possible. For instance, the result (\ref{95}) is also compatible with $\b_0=-1/\sqrt 8\simeq -0.3535$, describing the isotropic expansion of $d=8$ spatial dimensions! Incidentally, the number (and the kinematics) of the extra spatial dimensions play a crucial role also in the possible production of primordial ``seeds" for the large-scale magnetic fields \cite{mf}. 

It should be mentioned, finally, a possible non-Gaussian ``contamination" of the statistical properties of the anisotropy spectrum, possibly present in curvaton models with $\Om_\sg \ll1$ \cite{39} (see Eq. (\ref{94a})). A possible detection of non-Gaussianity, in future CMB measurements, could provide support to the curvaton mechanism, and could be used for a direct discrimination between this scenario and other, more standard scenarios based on slow-roll inflation.

\section{The relic dilaton background}
\label{sec2}

The accelerated evolution of the Universe, during the phase of pre-big bang inflation, amplifies the quantum fluctuations of all fields present in the string effective action: thus, in particular, it amplifies the dilaton fluctuations, $\da \phi \equiv \chi$.  The formation of a stochastic background of relic gravitational waves, associated to the amplification of the tensor part of metric fluctuations, is thus accompanied by the simultaneous formation of a comic background of relic dilatons \cite{41}, whose primordial (high-energy) spectral distribution tends to follow that of tensor metric perturbations \cite{12a}.

There is, however, a possible  important difference in the present intensity of the two cosmic backgrounds, due to the fact that dilatons -- unlike gravitons -- could become massive in the course of the standard (post-inflationary) evolution. Actually, dilatons {\em must} become massive if they are non-universally coupled to ordinary matter with gravitational strength (or higher) \cite{42,43}, to avoid the presence of  long-range scalar forces which are excluded by the standard gravitational phenomenology (in particular, by the high-precision tests of the equivalence principle). The induced mass may drastically modify the amplitude and the slope of the dilaton spectrum, in the frequency band associated to its  non-relativistic sector. 

For a simple illustration of the effects of the  mass on the spectrum we will consider here the model of vacuum, dilaton-dominated pre-big bang background described by Eq. (\ref{57}), smoothly joined at $\eta=-\eta_1<0$ to the standard radiation-dominated background with frozen dilaton, described by Eq. (\ref{52})  (we shall work in $d=3$ spatial dimensions). Perturbing the background equations \cite{12a} one finds, in this case, that the dilaton pump field is the same field $z_h\sim a \exp(-\phi/2)$ governing the amplification of metric fluctuations. Taking into account a possible mass contribution, $m^2= \pa^2 V/ \pa \phi^2$, one then  obtains for the Fourier modes $\chi_k$ the canonical equation:
\beq
\left( \chi z_h\right)_k''+ \left(k^2 + m^2 a^2 -{z_h''\over z_h} \right) \left( \chi z_h\right)_k=0.
\label{97}
\eeq

During the initial pre-big bang regime the potential is  negligible  ($m^2=0$), and the canonically normalized solution for $\chi_k$ is  that of Eq. (\ref{84}) (with $\nu_\sg$ replaced by $\nu_h$). In the subsequent radiation-dominated era $\phi$ stabilizes to a constant, so that $z_h \sim a \sim \eta$ and the effective potential $z_h''/z_h$ is vanishing. Assuming that the dilaton mass is small enough in string units, and considering the high-frequency sector of the spectrum,  associated to the relativistic modes of proper momentum $p=(k/a) \gg m$, we can  neglect also the mass term of Eq. (\ref{97}), to obtain the general solution
\beq
\chi_k= {1\over a\sqrt{2k}} \left[c_+(k) \e^{-i k\eta} + 
c_-(k) \e^{i k\eta}\right], ~~~~~~~~~~~~~ \eta \geq - \eta_1.
\label{98}
\eeq
Matching  $\chi$ and $\chi'$ with the pre-big bang solution (\ref{84}) at $\eta_1$, for super-horizon modes with $(k\eta_1)\ll1$, we are lead to
\beq
c_\pm(k)=\pm c(k) \e^{\mp i k\eta_1}, ~~~~~~~~~~~~~~~~~~
|c(k)| \sim (k\eta_1)^{-\nu_h -1/2}
\label{99}
\eeq
(modulo numerical factors with modulus of order one). Thus, at large times $\eta \gg \eta_1$,  
\beq
\chi_k \sim {c(k)\over a \sqrt k} \sin k \eta.
\label{100}
\eeq

The spectral energy density for the relativistic sector of the dilaton background, in the radiation era, in then determined by
\bea
k {\d \r \over \d k} &=&{k^3 \over 2 a^2} \left( |X_k'|^2 + k^2 |X_k|^2 \right)
\nonumber \\ 
&\sim & \left(k\over a\right)^4 |c(k)|^2 \sim
\left(k\over a\right)^4\left(k \over k_1\right)^{-2 \nu_h -1} = p^4 
\left(p \over p_1\right)^{-2 \nu_h -1},
\label{101}
\eea
where $k_1 \sim \eta^{-1}$ is the high-frequency cut-off scale. In units of critical energy density, $\r_c = 3 \Mp^2 H^2$,
\beq
\Om_\chi(p,t)= {p\over \r_c} {\d \r_\chi\over \d p} \sim
\left(H_1 \over\Mp\right)^2 \left(H_1 \over H\right)^2 
\left(a_1 \over a\right)^4 \left(p \over p_1\right)^\da,
~~~~ m < p < p_1,
\label{102}
\eeq
where we have defined the (model-dependent) slope parameter $\da =3-2 \nu_h>0$, and we have introduced the (time-dependent) proper momentum associated to the cut-off scale, $p_1= k_1/a= H_1a_1/a$, determined by the background curvature scale $H_1$ at the end of  inflation. In general, $(H_1/H)^2(a_1/a)^4 \equiv \r_r(t)/\r_c(t)\equiv \Om_r(t)$, and we may thus conclude that the relativistic sector of the dilaton spectrum, in the radiation era, is exactly the same as the spectrum of tensor metric perturbations (see Eq. (\ref{69})), in the same model of background. 

However, even if the  mass is small, and initially negligible,  the proper momentum $p=k/a(t)$ is continuously red-shifted with respect to $m$ during the subsequent cosmological evolution, so that all modes tend to become non-relativistic, $p <m$. For non-relativistic modes the solution (\ref{98}) is no longer valid, and the correct spectrum must refer to the exact solutions of Eq. (\ref{97}) with $m \not=0$. In the radiation era such a solution can be given in terms of the Weber cylinder functions \cite{44}, and one finds that the non-relativistic sector of the spectrum splits into two branches, with different slopes: a first branch of modes becoming non-relativistic at a time scale $t_{nr}$ when they are {\em already inside} the horizon,  with proper momentum $p$ such that $p(t_{nr}) \sim m \gg H(t_{nr})$; and a second branch of modes becoming non-relativistic when they are {\em still outside} the horizon, with $p(t_{nr}) \sim m \ll H(t_{nr})$. The two branches are separated by the momentum scale $p_m$ of the mode becoming non-relativistic just at the time of horizon crossing, i.e. $p(t_{nr}) = m = H(t_{nr})$, and thus related to the cut-of scale $p_1$ by
\beq
{p_m\over p_1}= {m\, a_{\rm nr}\over H_1a_1}= {m\over H_1} \left(H_1\over H_{\rm nr}\right)^{1/2} = \left(m \over H_1\right)^{1/2}.
\label{103}
\eeq

Without applying to the explicit form of the massive solutions of Eq. (\ref{97}), a quick estimate of the non-relativistic spectrum can be obtained \cite{45} by noting that, if $p_{nr}>H_(t_{nr})$, the number of produced dilatons is the same as in the relativistic case, and the only effect of the non-relativistic transition is a rescaling of the energy density, i.e.
\beq
\Om_\chi^{\rm rel} \ra 
\Om_\chi^{\rm nr} = \left(m \over p\right) \Om_\chi^{\rm rel}. 
\label{104}
\eeq
For this branch of the spectrum we then obtain, from Eq. (\ref{102}),
\beq
\Om_\chi(p,t) \sim \left(m\over H_1\right)
\left(H_1\over \Mp\right)^2
 \left(H_1 \over H\right)^2 
\left(a_1 \over a\right)^3 \left(p \over p_1\right)^{\da-1}, 
~~~ p_m < p < m.
\label{105}
\eeq

In the case $p_{nr}<H_{nr}$, on the contrary, the slope of the spectrum -- determined by the background kinematics at the time of horizon exit -- has to be the same as that of the relativistic sector, while the time-dependence has to be the non-relativistic one ($\r_\chi \sim a^{-3}$) of Eq. (\ref{105}). Continuity with the branch (\ref{105}) at $p=p_m$ then gives
\beq
\Om_\chi(p,t)\sim \left(m\over H_1\right)^{1/2}
 \left(H_1 \over \Mp\right)^2 
 \left(H_1 \over H\right)^2 
\left(a_1 \over a\right)^3 \left(p \over p_1\right)^{\da}, 
~~~ p_{\rm eq} <p< p_m.
\label{106}
\eeq
The lower limit $p_{\rm eq}<p$ has been inserted here to recall that we are neglecting the effects of the transition to the matter-dominate phase, i.e. we are considering modes re-entering the horizon during the radiation era, with $p>p_{\rm eq}=H_{\rm eq} \sim 10^{-27}$ eV. We should recall, also, that the spectrum has been computed in a radiation-dominated background, and  thus is valid, strictly speaking, only for $t>t_{\rm eq}$. 

The three branches (\ref{102}), (\ref{105}), (\ref{106}) describe the 
spectrum (between $p_{\rm eq}$ and $p_1$) of primordial dilatons   produced in the simple example of ``minimal" pre-big bang model that we have considered. We refer to the literature  for a more detailed computation, for a discussion of its transmission to the present epoch $t_0$, and for the possible modifications induced by generalized background evolutions (see e.g. \cite{26}). For the pedagogical purpose of this paper this example provides a sufficiently clear illustration of the  effects of the mass on the spectrum: in particular, it clearly illustrates  the enhancement produced at lower frequencies 
because of the reduced spectral slope of the branch (\ref{105}), which may  become even decreasing if $\da <1$ (see Fig. \ref{f5}). 

In such a context one is naturally lead to investigate whether this enhanced intensity might favor the detection of a non-relativistic dilaton background, with respect to other, relativistic types of cosmic radiation (such as the relic graviton background). 

\subsection{Light but non-relativistic dilatons}
\label{sec21}

For a phenomenological discussion of this possibility we must start with two important assumptions. The first is that the produced dilaton are light enough to have survived until the present epoch. Supposing that massive dilatons have dominant decay mode into radiation (e.g., two photons), with gravitational coupling strength, i.e. with a decay rate $\Ga \sim \lp^2 m^3$, it follows that the primordial graviton background is still  ``alive"  in the present Universe (characterized by the time-scale $H_0^{-1}$) provided $H_0^{-1} < \Ga^{-1}$, i.e. 
\beq
m \laq 10^2\, {\rm MeV}.
\label{107}
\eeq

The second assumption we need is that the total energy density of the dilaton background, integrated over all modes, turns out to be dominated by its non-relativistic sector. Only in this case we can evade the stringent bound imposed by the nucleosynthesis, which applies to the relativistic part of any cosmic background of primordial origin. 

The energy density of a relativistic background, in fact, evolves in time like the radiation energy density, $\r^{rel}/\r_{rad}= \Om^{rel}/\Om_{rad} =$ const: the present value of their ratio is thus the same as  the value of the ratio at the nucleosynthesis epoch. To avoid disturbing the nuclear processes occurring at that epoch, on the other hand, one must require that 
$\Om^{rel}/\Om_{rad} \laq 0.1$ \cite{46}. Using the present value of $\Om_{rad}$ one is lead then to the constraint $\Om^{rel}(t_0) \laq 5 \times10^{-6}$, which imposes a severe constraint on all relativistic primordial backgrounds. In particular, it imposes an upper limit on the peak value of the graviton background produced in models of pre-big bang inflation, thus determining the  minimal level of sensitivity required for its detection \cite{19}. 

The energy density of a non-relativistic background, on the contrary, evolves like the dark-matter density, and grows in time with respect to the radiation background: $\Om^{nr}/\Om_{rad}\sim a$. As a consequence, the value of $\Om^{nr}$ can be very large today, even if negligible at the nucleosynthesis epoch. The only constraint we must  apply, in this case, is the critical density bound, 
\beq
h^2\Om_\chi(t)= h^2\int^{p_1} \d (\ln p)\, \Om_\chi(p,t) < 1,
\label{108}
\eeq
to be imposed at any time $t$, to avoid a Universe over-dominated by such a cosmic background of dust matter. Here $h\simeq 0.73$ is the present value of the Hubble parameter $H_0$ in units of  $100$ km s$^{-1}$ Mpc. 

For the dilaton spectrum of Eqs. (\ref{102})--(\ref{106}) there are, in particular,  two different cases in which the total energy density is dominated by the non-relativistic modes.  A first (obvious) possibility is the case in which all modes of the spectrum are presently non-relativistic, namely $p_1(t_0)<m$ (in this case the branch (\ref{105}) extends from $p_m$ to $p_1$). This implies, however, that 
\bea
m> {H_1 a_1 \over a_0}&=&H_1{a_1\over a_{\rm eq}}{a_{\rm eq}\over a_0}
=H_1\left(H_{\rm eq}\over H_1\right)^{1/2}\left(H_{0}\over H_{\rm eq}\right)^{2/3} \nonumber \\ &\simeq& \left(H_1\over \Mp\right)^{1/2} 10^{-4} {\rm eV}. 
\label{109}
\eea
For a typical string-inflation scale, $H_1 \sim \Ms$, we obtain a lower limit on $m$ which is well compatible with the upper limit (\ref{107}), but which requires mass values too high to be compatible with the sensitivity band of present gravitational antennas 
(see Subsection \ref{sec22}). 

\begin{figure}
\centering
\includegraphics[height=5cm]{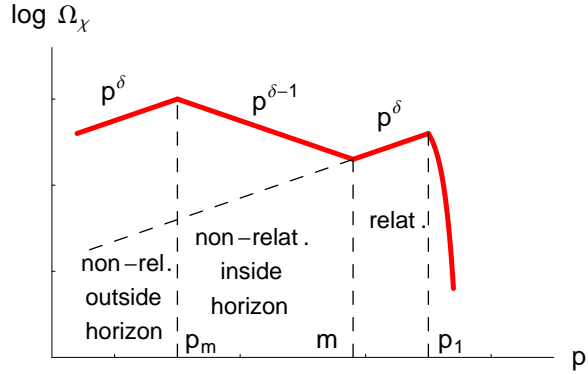}
\caption{Example of dilaton spectrum dominated by the non-relativistic sector. The spectrum is peaked at $p=p_m$, and the slope parameter satisfies the condition $\da<1$.}
\label{f5}      
\end{figure}

The second (more interesting) possibility is the case in which $m<p_1(t_0)$, but the parameter $\da$ is smaller than one, and the slope is flat enough, so that the spectrum is peaked not at $p_1$ but at $p_m= p_1(m/H_1)^{1/2}$ (see Fig. \ref{f5}). In that  case the momentum integral (\ref{108}) is dominated by the peak value $\Om_\chi(p_m)$, and the critical density bound can be approximated by the condition $\Om_\chi(p_m,t_0) \laq 1$. Using Eq. (\ref{105}), and  noting that in the matter-dominated era ($t>t_{\rm eq}$) the value of the non-relativistic spectrum keeps frozen at the equality value $\Om_\chi(t_{\rm eq})$, we are lead to the condition $\Om_\chi(t_{\rm eq},p_m) \laq 1$, which implies 
\beq
m \laq \left(H_{\rm eq} \,\Mp^4 \,H_1^{\da-4}\right)^{1/(\da+1)}.
\label{110}
\eeq
For $H_1 \sim \Ms$, and $\da \ra 0$, this bound can be saturated by masses as small as
\beq
m \sim H_{\rm eq} \left(\Mp\over \Ms\right)^4 \sim 10^{-23}\,{\rm eV}.
\label{111}
\eeq

It is quite possible, therefore, to have a dilaton mass small enough to  fall within the sensitivity range of present gravitational detectors, even if the energy density of the dilaton background is dominated by non-relativistic modes (thus evading the relativistic upper bound $\Om^{rel} \laq 10^{-6}$), and even if the background intensity is  large enough to saturate the critical density bound, $\Om_\chi \sim 1$. 

So small mass values, however, are necessarily associated with long-range dilaton forces: in particular, if the mass satisfies the condition $m<p_1(t_0) \sim (\Ms/\Mp)^{1/2} 10^{-4}$ eV (as in the example illustrated in Fig. \ref{f5}), the corresponding force has a range exceeding the centimeter. This might imply macroscopic violations of the equivalence principle (due to the non-universality of the dilaton coupling \cite{42}), and macroscopic deviations from the standard Newtonian form of the low-energy gravitational interactions (which seem to be excluded, however, by present experimental results \cite{47,48}). 

We should recall, in fact, that in the presence of long-range dilaton fields the motion of a macroscopic test body with nonzero dilaton charge is no longer described by a geodesics. There are forces on the test body due to the gradients of the dilaton field, according to the generalized conservation equation
\beq
\nabla_\nu T_\mu\,^{\nu}= {\sg \over 2} \nabla_\mu \phi,
\label{112}
\eeq
following from the application of the contracted Bianchi identity to the gravi-dilaton equations (\ref{3}) and (\ref{5}). The integration of this conservation equation over a (space-like) $t=$ const hypersurface then gives, in the point-particle (or monopole) approximation, the non-geodesic equation of motion \cite{49}
\beq
{\d u^\mu\over \d \tau} + \Ga_{\a \b}\,^\mu u^\a u^\b= q \nabla^\mu \phi,
\label{113}
\eeq
where $q$ is a dimensionless ratio representing the relative intensity of scalar to tensor forces (i.e., the effective dilaton charge per unit of gravitational mass of the test body). 

For the fundamental components of macroscopic matter, such as quark and lepton fields, the value of $q$ (or of the charge density $\sg$) is to be determined from an effective action which includes all relevant dilaton loop corrections \cite{42,12a}, and which is of the form
\bea
S= {1\over 2 \ls^2}\int \d^4 x \sqrt{-g} \Bigg[&- & Z_R(\phi) R-Z_{\phi}(\phi) (\nabla \phi)^2 - V(\phi) 
\nonumber\\ 
&+&Z^i_k(\phi) (\nabla \psi_i)^2- M_i^2 Z^i_m(\phi) \psi_i^2 \Bigg].
\label{114}
\eea
Here we have used, for simplicity, a scalar model of matter fields $\psi_i$, and we have called $Z$ the dilaton ``form factors" arising from  the loop corrections. The effective dilaton charge, therefore, turns out to be frame-dependent (the charge $q$ appearing in Eq. (\ref{113}), for instance, is referred to the S-frame action and to the S-frame equations (\ref{112})). The reason of such a frame dependence is that, in a generic frame, the metric and the dilaton fields are non-trivially mixed through the $Z_R$ and $Z_\phi$ coupling functions, so that the associated dilaton charge actually controls the matter coupling {\em not to the pure scalar part}, but {\em to a mixture of scalar and tensor part} of the gravi-dilaton field. 

A frame-independent and unambiguous definition of the dilaton coupling strengths can be given, however, in the canonically rescaled Einstein frame (E-frame), where the full kinetic part of the action (\ref{114}) (including the matter and gravi-dilaton sector) is diagonalized in terms of the canonically normalized fields $\wh g_{\mu\nu}$, $\wh \phi$ and $\wh \psi_i$ \cite{12a}. Assuming that the dilaton is stabilized by its potential, and expanding the Lagrangian term describing the interaction between $\wh \phi$ and $\wh \psi_i$ around the value $\phi_0$ which extremizes the potential, we can define, in this rescaled frame, the effective masses $\wh m_i$ and charges $\wh q_i$ for the canonical fields $\wh \psi_i$. In the weak coupling limit in which $Z_R \simeq Z_\phi \simeq \exp(-\phi)$ one then finds, in particular, that the canonical dilaton charge $\wh q_i$ deviates from the standard ``gravitational charge" by the dimensionless factor \cite{12a}
\beq
\overline q_i \equiv {\wh q_i \over  \sqrt{ 4 \pi G}\, \wh m_i} \simeq
1+ \left[ {\pa \over \pa \phi} \ln \left(Z^i_m \over Z^i_k \right)
\right]_{ \phi=\phi_0}.
\label{115}
\eeq

For a pure Brans-Dicke model of scalar-tensor gravity one has, for instance, $\overline q_i=1$ (because there is no dilaton coupling to the matter fields in the Jordan frame, where $\pa Z^i/\pa \phi=0$). For a string model, on the contrary, the coupling parameters  $\overline q_i$  deviate from $1$ and are non-universal,  in general, since the loop form factors $Z^i$  tend to be different for different fields $\psi_i$. In particular, in the conventional scenario which assumes that the loop corrections determining the coupling are the same determining also the effective mass of the given particle, one obtains  large dilaton charges ($\overline q_i \sim 50$) for the confinement-generated components of the hadronic masses \cite{42,43}, and  smaller charges ($\overline q_i \sim 1$) for the leptonic components. In that case, the total  dilaton charge of a macroscopic body tends to be large (in gravitational units) and composition-dependent \cite{49}, so that  a large dilaton mass ($m \gaq 10^{-4}$ eV) is required to avoid conflicting with known gravitational phenomenology. 

This conclusion can be avoided  if the loop corrections combine to produce a cancellation, in such a way that the value of the coupling parameters $\overline q_i$  turns out to be highly suppressed with respect to the  natural value of order one (a scenario of this type has been proposed, for instance, in \cite{50}). In that case $\overline q_i \ll1$, and light dilaton masses (as required, for instance, for a resonant interaction with gravitational antennas) may be allowed, without clashing with experimental observations. 

In the rest of this section we will focus our attention on this possibility, considering the response of the gravitational detectors to a cosmic background of massive, non-relativistic dilatons, assuming that the background energy density corresponds to large fraction of critical density, and that the dilatons are arbitrarily light and very weakly coupled to ordinary matter. 

\subsection{Dilaton signals in gravitational antennas}
\label{sec22}

The operation mechanism of all gravitational antennnas is based on the so-called equation of ``geodesic deviation" (see e.g. \cite{51}), which governs the response of the detector to the incident radiation. Such an equation is obtained by computing the relative acceleration between the world-lines of two nearby test particles, separated by the infinitesimal space-like vector $\eta^\mu$, and evolving geodesically in the given gravitational background. The interaction
with a dilaton  background can be easily included, in this context,  by replacing the  geodesic paths of the test particles with the world-lines described by Eq. (\ref{113}): one is lead, in this way, to a generalized equation of deviation \cite{49},
\beq
{D^2 \eta^\mu \over D \tau^2}+ R_{\nu\a\b}\,^\mu \eta^\nu u^\a u^\b = q \,\eta^\nu \nabla_\nu \nabla^\mu \phi,
\label{116}
\eeq
which is at the ground of the response of a detector to a background of gravi-dilaton radiation (the symbol $D$ denotes covariant differentiation along a curve parametrized by the affine time-like variable $\tau$).  

This equation implies that a gravitational detector can interact with the scalar radiation in two ways: either
\begin{itemize}
\item[$i)$] {\em directly}, through the {\em non-geodesic} coupling of its scalar charge to the second derivatives of the scalar background \cite{49,52}; or 
\item[$ii)$]  {\em indirectly}, through the {\em geodesic} coupling of its gravitational charge to the {\em scalar part} of the metric fluctuations induced by the dilaton, and contained inside the Riemann tensor \cite{53}. 
\end{itemize}

For a precise discussion of the response of the detector we need to compute the ``physical strain" $h(t)$ induced by the scalar radiation, which is expressed in terms of the so-called ``antenna pattern functions" $F(\theta, \phi)$, describing the detector sensitivity along the different angular directions. To this purpose, we shall rewrite Eq. (\ref{116}) in the approximation of small displacements $\xi^\mu$ around the unperturbed path of the text bodies, by setting $\eta^\mu= L^\mu+ \xi^\mu(\tau)$, with $L^\mu=$ const. We then obtain, in the non-relativistic limit,
\beq
\ddot \xi^i= -L^k M_k\,^i,  
\label{117}
\eeq
where
\beq
M_k\,^i= R_{k00}\,^i + q  \pa_k \pa^i \phi 
\label{118}
\eeq
is the total (scalar-tensor) stress tensor describing the ``tidal" forces due to the incident radiation. For the pedagogical purpose of this paper we shall assume that the tensor (i.e., gravity-wave) part of the radiation is absent, and that the scalar radiation can be simply described as a linear fluctuation of the Minkowski metric background $\eta_{\mu\nu}$ and of a constant dilaton background $\phi_0$: thus, in the longitudinal gauge,
\bea
&&
ds^2= \left(\eta_{\mu\nu} + \da g_{\mu\nu}\right) dx^\mu dx^\nu= (1+2 \psi) dt^2- (1-2 \vp)\da_{ij} dx^idx^j,
\nonumber \\ &&
\phi=\phi_0 + \chi,
\label{119}
\eea
so that 
\beq
M_{ij}=  \pa_i\pa_j \vp - \da_{ij} \ddot \psi- q \pa_i \pa_j \chi.
\label{120}
\eeq

To discuss the detection of a stochastic background of massive scalar radiation it is also convenient to expand the fluctuations in Fourier modes of proper momentum $\vec p= p \wh n$ and frequency $\nu= E(p)=(p^2+m^2)^{1/2}$, where the unit vector $\wh n$ specifies the propagation direction of the given mode on the angular two sphere $\Om_2$. We obtain
\bea
&&
M_{ij}={1\over 2} \int_{-\infty}^{\infty} \d p \int_{\Om_2} \d^2 \wh n \,
(2 \pi E)^2 \Bigg[ \da_{ij} \psi(p, \wh n) -  n_i  n_j \vp (p, \wh n)
+{m^2\over E^2}  n_i  n_j  \vp (p, \wh n)
\nonumber \\ &&
~~~~~~~~
+ q {p^2 \over E^2}  n_i   n_j  \chi(p, \wh n) \Bigg] \, \e^{2\pi i ( p \wh n \cdot \vec x -Et)}  + {\rm h.c.}
\label{121}
\eea
(note that we are using  ``unconventional" units in which $h=1$, i.e. $\hbar=1/2\pi$, for an easier comparison with the experimental variables). We will also assume that the dilaton is the only source of scalar metric perturbations, so that $\vp=\psi$ \cite{34}). Introducing the transverse and longitudinal projectors of the scalar stresses, defined respectively by
\beq
T_{ij}=\da_{ij}-   n_i  n_j, ~~~~~~~~~~~~~~~~
L_{ij}=   n_i   n_j,
\label{122}
\eeq
defining $M_{ij}= -\ddot F_{ij}$, and projecting the stress tensor onto the detector tensor $D^{ij}$ (specifying the geometric configuration and the orientation of the arms of the detector), we finally obtain the scalar strain as \cite{52,54,55}
\bea
h(t) \equiv D^{ij} F_{ij}&=&
{1\over 2} \int_{-\infty}^{\infty} \d p \int_{\Om_2} \d^2 \wh n \,
\Bigg[ F^{\rm geo}(\wh n) \psi (p, \wh n)
\nonumber \\ &+&
F^{\rm ng} (\wh n) \chi(p, \wh n)\Bigg]
 \e^{2\pi i ( p \wh n \cdot \vec x -Et)} 
+ {\rm h.c.}.
\label{123}
\eea
Here
\bea
&&
F^{\rm geo}= D^{ij}\left(T_{ij}+ { m^2\over E^2}L_{ij}\right),
\label{124} \\ &&
F^{\rm ng}=  q {p^2 \over E^2}D^{ij}  L_{ij},
\label{125}
\eea
are the antenna pattern functions corresponding, respectively, to the geodesic (or indirect) and non-geodesic (or direct) interaction of the detector with the scalar radiation background. 

It should be noted that the scalar radiation, differently from the case of the tensor component, contributes to the response of the detector also with its longitudinal polarization states. The longitudinal contribution is  present also in the ultra-relativistic limit $m \ra 0$, $ p \ra E$, thanks to the non-geodesic coupling (\ref{125}). In the opposite, non-relativistic limit $p \ra 0$, $ E \ra m$, the geodesic strain tends to become isotropic, $T_{ij} + ( m/E)^2 L_{ij} \ra \da_{ij}$, while the non-geodesic one becomes sub-leading. 

The results (\ref{123}) is valid for any type of  detector described by the response tensor $D^{ij}$, and is formally similar to the expression for the strain obtained in the case of tensor gravitational radiation -- modulo the presence of different pattern functions, due to the different polarization properties. The scalar strain (\ref{123}) can thus be processed, following the standard procedure, to correlate the outputs of two detectors and to extract the so-called signal-to-noise ratio (SNR), representing the experimentally relevant variable for the detection of a stochastic background of cosmic radiation \cite{56}. 

For our scalar massive background, with spectral energy density $\Om(p)$, we obtain \cite{52,54,55}, in particular,
\bea
&&
SNR=
{3NH_0^2\over 8\pi^3}\Bigg[2T \int_{0}^{\infty}{\d p \over  p^3\,(p^2+{ m}^2)^{3/2}} 
\frac{\ga^2(p)\,\Om^2(p)}
{P_1(\sqrt{p^2+{ m}^2})\,P_2(\sqrt{p^2+{ m}^2})}\Bigg]^{1/2} 
\nonumber \\ &&
\label{126}
\eea
(see also \cite{26} for a detailed computation). Here $T$ is the total (experimental) correlation time, $N$ an (irrelevant) normalization factor, $P_1$ and $P_2$ the noise power spectra of the two detectors, and $\ga(p)$ the so-called ``overlap reduction function", which modulates the correlated signal according to the relative orientation and distance of the detectors, located at the positions $\vec x_1$ and $\vec x_2$ :
\beq
 \ga(p)= {1\over N} 
 \int_{\Om_2} \d^2 \wh n~F_{1 }(\wh n)~ F_{2}(\wh n) 
\,\e^{2 \pi i p \wh n \cdot(\vec x_1-\vec x_2)}.
\label{127}
\eeq
The overlap is to be calculated with the geodesic pattern function $F_i^{geo}$ of Eq. (\ref{124}) if we are considering the indirect signal due to a spectrum of scalar metric fluctuations,  $\Om_\psi(p)$; it is to be calculated with the non-geodesic pattern function $F_i^{ng}$ of Eq. (\ref{125}) if we are considering, instead, the direct signal due to a spectrum of dilaton fluctuations, $\Om_\chi(p)$. 

We are now in the position of stressing another important difference from the case of pure tensor radiation, due to the presence of the mass in the noise power spectra $P_i$. For a typical power spectrum, in fact, the minimum level of noise is reached around a rather narrow frequency band $\nu_0$: outside that band the noise rapidly diverges,  and the signal (\ref{126}) tends to zero. As $\nu= (p^2 + m^2)^{1/2}$ we have, in principle, three possibilities.
\begin{itemize}
\item[$1)$] If $ m \gg \nu_0$ then the noise is always outside the sensitivity band $P_i(\nu_0)$, and the signal is always negligible.
\item[$2)$] If $ m \ll \nu_0$ then the sensitivity band 
may only overlap with the relativistic sector of the spectrum, for  $p \sim \nu \sim \nu_0$. 
\item[$3)$] If $ m \sim \nu_0$, finally, the whole non-relativistic part of the spectrum  $p \laq  m$ satisfies the condition  $P_i(\nu) \sim P_i( m) \sim P_i(\nu_0)$.
\end{itemize}
It is thus possible to obtain a resonant response to a massive, non-relativistic background of scalar particles, provided the mass lies in the band of maximal sensitivity of the two detectors \cite{52,54}. Considering the present, Earth-based gravitational antennas, operating between the  Hz and the kHz range, it follows that the maximal sensitivity is presently in the mass range
\beq
10^{-15} \,{\rm eV} \laq m \laq 10^{-12}\, {\rm eV}.
\label{128}
\eeq

Amusingly enough, it turns out that such small values are not so unrealistic if the dilaton mass is perturbatively generated by the  mechanism of radiative corrections. For a scalar particle,  gravitationally coupled  to fermions of mass $M_f$ with dimensionless strength $q,$  there are, in fact, quantum loop corrections to the mass of order $q M_f(\La/\Mp)$, where $\La$ is the cut-off, which we shall assume typically localized at the  TeV scale (see for instance \cite{57}). Considering the dilaton coupling to ordinary baryonic matter ($M_f \sim 1$ GeV) the induced mass is then:
\beq
m \sim q \left(\La\over 1\, {\rm Tev}\right) \left(M_f \over 1\, {\rm Gev}\right) \times 10^{-6} \, {\rm eV}.
\label{129}
\eeq
Thus, a value of $q$ smaller than (but not very far from) the present upper limits \cite{47} (imposing $q\laq 10^{-4}$ in the relevant mass range (\ref{128})) is perfectly compatible with the possibility of resonant response of the present detectors. 

Quite independently from the possible origin of the dilaton mass, if we assume that the mass is in the resonant range (\ref{128}), and that the bounds on $q$ are satisfied, we find that a cosmic background of non-relativistic dilatons is possibly detectable by the interferometric antennas of second generation -- such as Advanced and Enhanced LIGO -- provided the background energy density is sufficiently close  to the saturation of the critical density bound \cite{52,54}. This interesting possibility can be illustrated by considering, for an approximate estimate, the simplified situation of two identical detectors with $P_1=P_2=P$, responding non-geodesically with maximal allowed overlap $N \ga^{ng} \simeq q^2(4 \pi/15)$ (the numerical factor is referred to the particular case of  interferometric antennas). Let us  suppose, also, that the SNR integral (\ref{126}) is dominated by the peak value $\Om_m$ of the non-relativistic dilaton spectrum, and that such value is reached around $p=m$ (otherwise the response is suppressed by the factor $(p/m)^4$, \cite{54}). Eq. (\ref{126})  gives, in this case,
\beq
SNR \sim 
{\sqrt{2T}\over 10 \pi^2} {q^2 H_0^2\Om_m \over m^{5/2} P(m)},
\label{130}
\eeq
and the condition of detactable background (SNR $\gaq 1$) implies
\beq
m^{5/2} P(m) \laq  q^2 h^2 \Om_m
\left(T\over 4 \times 10^7 \, {\rm s}\right)^{1/2}\times 10^{-33}\, {\rm Hz}^{3/2}  .
\label{131}
\eeq

\begin{figure}
\centering
\includegraphics[height=5cm]{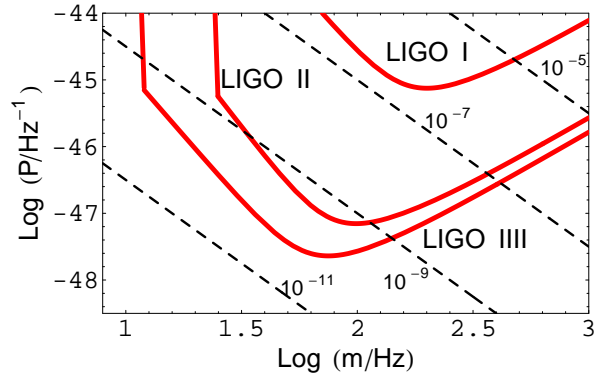}
\caption{The noise power spectra of the three LIGO generations (bold curves), and the condition of detectable dilaton background (dashed lines), plotted at different values of the parameter $q^2 h^2 \Om_m$ (ranging from $10^{-5}$ to $10^{-11}$).}
\label{f6}      
\end{figure}

This condition is compared in Fig. \ref{f6} with the expected spectral noise of the three LIGO generations (see e.g. \cite{58}), for $T= 4 \times 10^7$ s. The region of the plane $\{m,P\}$ corresponding to a detectable background is located  {\em above} the bold noise curves (labelled by LIGO I, LIGO II and LIGO III), and {\em below} the dashed lines, representing the upper limit (\ref{131}) for different constant  values of the parameter $q^2 h^2 \Om_m$. This  limit may be interpreted either as a constraint  on the intensity $\Om_m$, for  backgrounds geodesically coupled ($q^2=1$) to the detectors, or as a limit on the non-geodesic coupling strength $q^2$, for backgrounds of given energy density $\Om_m$. As shown in the picture, phenomenologically allowed backgrounds are in principle accessible to the sensitivity of next-generation interferometers (see also \cite{26} for a more detailed discussion). 

\subsection{Enhanced signals for flat non-relativistic spectra}
\label{sec23}

The result reported in  Eq. (\ref{130}) is generally valid for a growing   spectrum with a steep enough slope, as typically obtained in ``minimal" models of pre-big bang inflation. However, the cross-correlated signal may result  strongly enhanced with respect to Eq. (\ref{130}) if the dilaton spectrum is sufficiently flat,  and if the considered pair of detectors satisfies the condition $\ga(p) \ra $ const $\not=0$ for $p \ra 0$. 

Let us consider, in fact, the SNR integral (\ref{126}), which can be written as 
\beq
(SNR)^2 \sim T \int_0^{p_1} \d p {\ga^2(p) \Om^2(p) \over p^3 E^3 P_1(E)P_2(E)},
\label{132}
\eeq
where $E= (p^2+m^2)^{1/2}$, and where we can assume that $\Om(p)$ is a power-law function of $p$, with an ultraviolet cut-off at $p=p_1$. For a massless spectrum ($p=E$) this integral is always convergent (for any slope), even in the infrared limit $p=E \ra 0$: in fact, when $p \ra 0$, the physical strains are produced outside the sensitivity band of the detectors, and the noises blow up to infinity, $P_i(E) \ra P_i(0) \ra \infty$. For $m \not=0$, on the contrary, in the infrared limit $ p \ra 0$  the noises keep frozen at the frequency scale determined by the mass of the scalar background, $P_i(E) \ra P_i(m)=$ const. In this second case the behavior of the integral dependes on $\ga(p)$ and $\Om(p)$. 

Suppose now that $\ga(p) \ra \ga_0=$ const for $p \ra 0$, and that $\Om(p) \sim p^\da$, for $p<m$. For $\da<1$ we find that the integral is dominated by the infrared limit, and gives 
\bea
(SNR)^2 &\sim& {T \ga_0^2 \over  m^3 P_1(m)P_2(m)} 
\int_0^{ m} {\d p \over p^3} \Om^2(p)
\nonumber \\ &=&
 {T \ga_0^2 \over  m^3 P_1P_2} \,\left[ p^{2 (\da-1)} \right]_0^m.
\label{133}
\eea
Thus, the integral  is infrared divergent \cite{59} for all spectra (even if blue, $\da>0$) with $\da<1$ ! 

This divergence is obviously unphysical, and can be removed by noting  that the observation time $T$ is finite, and is thus associated to a minimum resolvable frequency interval $\Da \nu= \Da E = \Da (p^2/2m)\gaq T^{-1}$, defining  the  minimum momentum scale 
\beq
 p_{\rm min} = \left(2  m/T\right)^{1/2}>0,
\label{134}
\eeq
acting as effective infrared cut-off for the integral (\ref{133}).  This implies a modified dependence of SNR on the correlation time $T$ in the case of flat enough spectra:
\beq
SNR \sim T^{1/2} \left[p^{\da-1}\right]_{p_{\rm min}}^{ m}
\sim
\left\{
\begin{array}{ll}
 T^{1/2}, ~~~~~~~~~~~~~\da>1,\\
T^{1-\da/2}, ~~~~~~~~~~ \da<1.	
\end{array}  
\right.
\label{135}
\eeq
For $\da <1$, in particular, there is a faster growth of SNR with $T$, which may produce an important enhancement of the sensitivity to a cosmic background of non-relativistic scalar particles, as discussed in \cite{55,59}. 

It is important to stress that the case $\ga(p) \ra \ga_0=$ const for $p \ra 0$ has not been ``invented" {\em ad hoc}: it can be implemented, in practice, with detectors  already existing and operative (or with detectors planned 
to be working in the near future, like resonant spheres). A first simple 
example, studied in \cite{59}, refers in fact to  spherical, resonant-mass detectors, whose monopole mode is characterized by the ``trivial" response tensor $D^{ij}= \da^{ij}$. In that case the geodesic pattern function (\ref{124}) is isotropic,
\beq
F^{\rm geo}= {2 p^2 + 3  m^2\over p^2 +  m^2}, 
\label{136}
\eeq
and the geodesic overlap function (\ref{127}), for two identical spheres with spatial separation $|\vec x_1- \vec x_2|=d$, is given by 
\beq
\ga(p)={2\over N} \left(2 p^2 + 3  m^2\over p^2 +  m^2  \right)^2\, {\sin (2 \pi p d)\over p d}. 
\label{137}
\eeq
This function clearly satisfies the requirement $\ga(p) \ra \ga_0=$ const for $p \ra 0$. 

A second example, studied in \cite{55}, refers to the so-called ``common mode" of the interferometric antennas, characterized by the 
 response tensor 
 \beq
 D_+^{ij}= u^iv^j+v^iu^j,
 \label{138}
 \eeq
 where $u^i$ and $v^i$ are the unit vectors specifying the spatial orientation of the axes of the interferometer. Let us consider, for instance,  a geometrical configuration where the vectors $\wh u$ and $\wh v$ are coaligned with the $x_1$ and $x_2$ axes of a Cartesian frame, respectively, and the direction $\wh n$ ofthe incident radiation is specified (with respect to the axes $x_1$, $x_2$ and $x_3$) by the polar and azimuthal angles $\vp$ and $\theta$. The computation of the geodesic pattern function (\ref{124}) gives, in that case,
\beq
F_{+}^{\rm geo}=2-\left(p\over E\right)^2 \sin^2 \theta .
\label{139}
\eeq
The geodesic overlap function (\ref{127}), for two coplanar interferometers with spatial separation $|\Da \vec x|=d$, is \cite{55}
\bea
&&
\ga_+^{\rm geo}(p)= {4 \pi \over N} \Bigg[\left(4-4{p^2\over E^2}+{p^4\over E^4}\right)j_0(\a)
+{1\over \a}\left(4{p^2\over E^2}-2{p^4\over E^4}\right)j_1(\a)
\nonumber \\&&
~~~~~~~~~~~~~~~~~
+{3\over \a^2}\left(p\over E\right)^4 j_2(\a)\Bigg],
\label{140}
\eea
where $\a=2 \pi p d$, and $j_0$, $j_1$, $j_2$ are spherical Bessel functions. Thus, also in this case, $\ga \ra 16 \pi/N=$ const for $p \ra 0$.

\section{Late-time cosmology: dilaton dark energy}
\label{sec3}

In this third lecture we will discuss the possibility that a homogeneous, large-scale dilaton field may be the source of the so-called ``dark energy" which produces the cosmic acceleration first observed at the end of the last century \cite{60}, and confirmed by most recent supernovae data \cite{61,61a}. 

Let us recall, to this purpose, that the initial phase of pre-big bang inflation is characterized by the monotonic growth of the dilaton and of the string coupling $g_s$ (see Sect. \ref{sec13}): the subsequent epoch of standard evolution thus opens up in the strong coupling regime, and should be described by an action which includes all relevant loop corrections. Late enough, i.e. at sufficiently low curvature scales,  the higher-derivative corrections can be neglected, and the action can be written in the form of Eq. (\ref{114}). In that context the loop form factors $Z(\phi)$, and the dilaton potential $V(\phi)$, may play a crucial role in determining the late-time cosmological evolution. 

There are, in principle, two possible alternative scenarios. 
\begin{itemize}
\item[$i)$]The dilaton is stabilized by the potential at a constant value $\phi=\phi_0$ which extremizes $V(\phi)$. In this case the loop corrections induce a constant renormalization of the effective dilaton couplings (as discussed in Sect. \ref{sec21}), and the Universe may approach a late-time configuration dominated by the dilaton potential,  with $H^2 \sim V(\phi_0)$. 
\item[$ii)$]The dilaton fails to be trapped in a minimum of the potential, and keeps running even during the post-big bang evolution. In this case the late-time cosmological evolution is crucially dependent on the asymptotic behavior of the factors $Z(\phi)$. 
\end{itemize}
These two different possibilities have different impact on the so-called ``coincidence problem" (i.e. on the problem of explaining why the dark-matter and dark-energy densities are of the same order just at the present epoch), as we shall discuss in the following subsections. 

\subsection{Frozen dilaton in the moderate coupling regime}
\label{sec31}

\begin{figure}
\centering
\includegraphics[height=5cm]{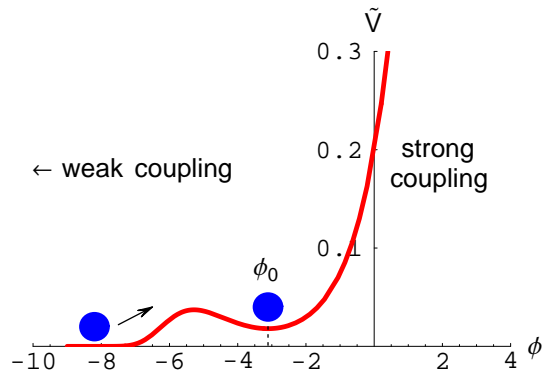}
\caption{Plot of the potential (\ref{141}) for $k_1=k_2=\b=\ga=1$, $\ep=0.1$, $\phi_1=-3$, $m_V=0.1$, and  a local minimum (independent of $m_V$) at $\phi_0= -3.112$, corresponding to $g_s^2= \exp(\phi_0) \simeq 0.045$.}
\label{f7}      
\end{figure}

The first type of scenario can be easily implemented \cite{62} using a generic non-perturbative potential which is instantonically suppressed ($V \sim \exp(-1/g_s^2)$) in the weak coupling limit $g_s^2 \ra 0$, and which develops a non-trivial structure with a (semi-perturbative) minimum $g_s^2 \sim \a_{GUT} \sim (\Ms/\Mp)^2 \sim 0.1-0.01$ in the regime of moderate string coupling. A typical example is the ``minimal" potential given, in the E-frame, by \cite{63}
\beq
\ti V(\phi)= m_V^2 \left[ \e^{k_1(\phi-\phi_1)}+ \b \e^{-k_2(\phi-\phi_1)}\right]
\e^{- \ep \exp \left[-\ga(\phi- \phi_1)\right]},
\label{141}
\eeq
where $k_1$, $k_2$, $\b$, $\ep$, $\ga$ are dimensionless parameters of order one (see Fig. \ref{f7}). The presence of  a local minimum at $\phi_0 \simeq \phi_1$ allows  solutions with $\phi=$ const during the radiation-dominated phase, and (for appropriate values of $m_V$)  may also lead to a late phase of accelerated expansion driven by the potential energy $V(\phi_0)$, {\em provided} the dilaton is not permanently shifted away from the minimum $\phi_0$ by the transition to the matter-dominated epoch  \cite{62}. 

Let us consider, in fact, the equation of motion of a homogeneous  dilaton field $\phi(t)$ in the conformally rescaled E-frame (with metric $\ti g$) where the graviton kinetic energy is canonically normalized, and let us assume that the rescaled matter sources can be described as a perfect fluid of energy density $\ti \rho$, pressure $\ti p$, and dilaton charge $\ti \sg$. Starting from an action of the type (\ref{114}) we find that the generalized dilaton equation, for a cosmological background, takes the form 
\beq
A(\phi) \left(\ddot \phi + 3 \ti H \fpu\right) + B(\phi) \fpu^2 +
{\pa \ti V\over \pa \phi} + \lp^2 \left[C(\phi) \left( \ti \r - 3 \ti p\right) + \ti \sg\right]=0,
\label{142}
\eeq
where $A$, $B$ and $C$ are functions describing the rescaled (E-frame) loop corrections.  For a minimally coupled field, for instance, $A=1$, $B=C=\ti \sg=0$; for the dilaton, at tree-level in the string coupling, $A=C=1$, $B=0$. In the most general case we find that a stable dilaton configuration with $\fpu=0=\fpp$ is  possible, in the radiation era  ($\ti \r=3\ti p$), if the scalar charge of the fluid is negligible, $\ti \sg=0$, and the dilaton extremizes the E-frame potential, $\pa \ti V/\pa \phi=0$. 

When the Universe becomes matter-dominated ($\ti p=0$), however, 
a new acceleration $\fpp= - A^{-1} \lp^2 C \ti \r$ is suddenly generated,  which tends to remove the dilaton away from its equilibrium position. Such an acceleration is in competition with the restoring force $\fpp=- A^{-1} (\pa \ti V/\pa \phi)$ (see Eq. (\ref{142})). The possibility that the dilaton may bounce back to the stable minimum $\phi=\phi_0$, driving the Universe towards a final phase of accelerated, potential-dominated expansion, thus crucially depends on the values of two parameters:  the (loop-corrected) strength $\lp^2 C(\phi_0)$ of the dilaton coupling to dark matter, and  the slope of the dilaton potential (\ref{141}), determined by the mass scale $m_V$ which also controls the amplitude of the minimum, $V(\phi_0) \sim m_V^2$. Such an amplitude, on the other hand, should correspond to the present Hubble scale ($V(\phi_0)\sim H_0^2$), in a realistic model able to describe the present phase of accelerated expansion. 

It can be shown, with a simple numerical analysis, that the values of the  coupling strength allowed by present gravitational phenomenology are compatible with a late-time phase dominated by the potential only for a finite range of values of $V(\phi_0)$, depending on the value of the dilaton coupling at the equality epoch \cite{62}. Using  
the phenomenological upper limit $|C_{\rm eq}| \simeq 0.1$ one finds that the dilaton, after a smal shift at $t=t_{\rm eq}$, bounces back to the minimum provided $10^{-7}H_{\rm eq} \laq m_V \laq H_{\rm eq}$ (which includes the realistic case $m_V \sim H_0 \sim 10^{-6}H_{\rm eq}$) \cite{62}. Smaller values of $|C_{\rm eq}|$ correspond to a larger mass interval. We can say, therefore, that the coincidence problem (i.e., why 
$V(\phi_0)\sim H_0^2$), in this context remains, but is somewhat alleviated because -- thanks to the dynamical correlation between the amplitude $V(\phi_0)$ and the matter-dilaton coupling -- only a restricted range of values is allowed for $V(\phi_0)$. 

\subsection{Running dilaton: saturation of the loop corrections and  asymptotic ``freezing"}
\label{sec32}

\begin{figure}
\centering
\includegraphics[height=5cm]{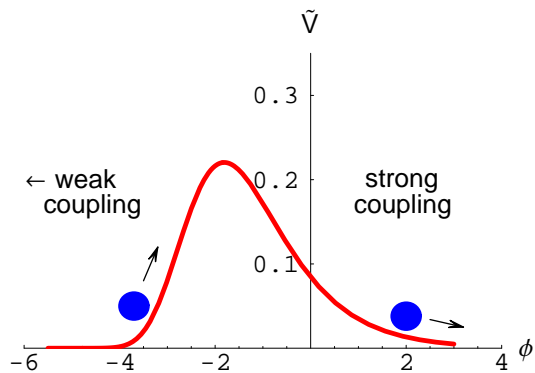}
\caption{Plot of the potential (\ref{143}) for $b_1=1$, $c_1=10$, 
$\b_1=0.1$, $\b_2=0.2$ and  $m_V=1$. The dilaton is monotonically growing from the string perturbative vacuum along a ``bell-like" non-perturbative potential.}
\label{f8}      
\end{figure}

The second possibility, which will be discussed here in more detail, in the case in which the dilaton is not stopped by the structures formed by the potential around $g_s^2=1$, and keeps rolling towards $+\infty$ along a smoothly decreasing potential. A possible example of non-perturbative potential of this type is given, in the E-frame, by \cite{64}
\beq
\ti V= c_1^4 m_V^2 \left( \e^\phi\over b_1+c_1^2 \e^\phi\right)^2 \left[ \e^{- \b_1 \exp(-\phi)}- \e^{- \b_2 \exp(-\phi)}\right],
\label{143}
\eeq
where $b_1$, $c_1$, $\b_1$, $\b_2$ are dimensionless parameters, with $0<\b_1<\b_2$. This potential is instantonically suppressed in the weak coupling limit $\phi \ra -\infty$, and is exponentially decaying as
\beq
 \ti V = m_V^2 \left(\b_2-\b_1\right) \e^{-\phi}+ {\cal O} \left(\e^{-2\phi}\right)
 \label{144}
 \eeq
 in the limit $\phi \ra +\infty$ (see Fig. \ref{f8}). In this case, as we shall see, we can obtain a scenario of ``coupled quintessence" \cite{65} in which the late Universe approaches a (possibly accelerated) state  dominated by a mixture of kinetic and potential energy density,  and the coincidence problem may find a satisfactory solution thanks to the  dilaton-dark matter interactions. 

In this case, however, a realistic scenario requires some mechanism of saturation of the loop corrections, so as to keep the present effective values of gravitational and gauge couplings approximately constant and sufficiently ``small", even in the large ``bare coupling" limit $\phi \ra +\infty$. As discussed in \cite{66}, such a saturation can be obtained thanks to the large number of fields (e.g. gauge bosons) entering the loop corrections, assuming (as in models of  ``induced gravity") that the loop form factors of Eq. (\ref{114}) have a finite limit for $\phi \ra +\infty$, and that can be approximated by a Taylor expansion in  powers of the inverse bare coupling $g_s^2 =\exp \phi$. Applying these assumptions to the gravi-dilaton form factors, to the potential, and to the dimensionless parameters $q_i(\phi)$ controlling the dilaton 
charge-density of the various matter fields, we can set, for $ \phi \ra +\infty$, 
\bea
&&
Z_R(\phi)= c_1^2 + b_1 \e^{-\phi} + {\cal O} (\e^{-2 \phi}),
\nonumber \\ &&
Z_\phi(\phi)= - c_2^2 +b_2 \e^{-\phi} + {\cal O} (\e^{-2 \phi}),
\nonumber \\ &&
 V(\phi)= V_0 \e^{-\phi} + {\cal O} (\e^{-2 \phi}),
\nonumber \\ &&
q_i(\phi)= q_{0i}+ {\cal O} (\e^{-2 \phi}) .
\label{145}
\eea

The dimensionless coefficients $c_1^2$ and $c_2^2$ of this expansion are typically of order $N \sim 10^{-2}$, because of their quantum-loop origin and of the large number $N$ of gauge bosons in GUT groups like $E_8$. This is in agreement with the fact that $c_1^2$ controls (according to the action (\ref{114})) the asymptotic value of the ratio between the string and the Planck length scale, $c_1^2=(\ls/\lp)^2$, which is indeed expected to be a number of the above order.    The coefficients $b_1, b_2 \dots$, on the contrary, are numbers of order one. Note  that the expansion of $V(\phi)$ agrees with the asymptotic form of the potential (\ref{144}). 

We should note, finally, that the asymptotic values of the dilaton charges, $q_{0i}$, have to be strongly suppressed for the ordinary components of matter (such as baryons) and for  electromagnetic radiation:  if we want a dilaton field active on a cosmological scale of distances, in fact,  we need 
long-range interactions, and we must avoid  unacceptable  deviations from the standard gravitational phenomenology by suppressing the dilaton couplings, as discussed in Sect. \ref{sec21}. 
For the (possibly exotic) components of dark matter, however, there is  no strict phenomenological bound imposing such suppression: in that case, the asymptotic charge $q_0$ could be nonvanishing, and of order one, leading to interesting late-time deviations from the standard cosmological scenario. 

For a simpler illustration of this possibility it is convenient to work in the diagonalized E-frame, obtained from the metric $g$ of Eq. (\ref{114}) through the rescaling
\beq
g_{\mu\nu}= c_1^2\, Z_R^{-1}\, \ti g_{\mu\nu} .
\label{146}
\eeq
The action (\ref{114}) becomes, in this new frame
\beq
S= {1\over 2 \lp^2} \int \d^4 x \sqrt{-g} \left[ - \ti R +{1\over 2} k^2(\phi) \left(\ti \na \phi\right)^2 - \ti V(\phi)\right] + S_m(\ti g, \phi, {\rm matter}),
\label{147}
\eeq
where
\beq
k^2(\phi)= 3 \left( \pa \ln Z_R\over \pa \phi \right)^2 - 2 {Z_\phi\over Z_R}, ~~~~~~~
\ti V(\phi)= c_1^4 Z_R^{-2}\,V.
\label{148a}
\eeq
Assuming that the matter action $S_m$ describes a perfect fluid  with  a dark-matter component $\ti \r_m$, a baryon component $\ti \r_b$, and a radiation component $\ti \r_r = 3 \ti p_r$, the cosmological Einstein equations for the action (\ref{147}) can then be written (omitting the tilde,  and in units $2 \lp^2=1$) as:
\bea
&&
6H^2 = \r_r+\r_b+ \r_m+\r_\phi,
\nonumber\\ &&
4 \dot H+ 6H^2= -{\r_r\over 3} - p_\phi,
\label{148}
\eea
where 
\beq
\r_\phi= {k^2(\phi)\over 2} \dot \phi^2 +V, ~~~~~~~~~~~~
p_\phi= {k^2(\phi)\over 2} \dot \phi^2 -V.
\label{149}
\eeq
The associated dilaton equation, assuming a negligible density of dilaton charge for baryons and radiation ($\sg_r=0=\sg_b$), can be written as \cite{64}
\beq
k^2( \ddot \phi+3H\dot \phi)+ k k' \fpu^2 + V'+{1\over 2} \left[ \psi' \left(\r_b+\r_m\right) +\sg_m\right]=0,
\label{150}
\eeq
where we have defined $\psi=- \ln Z_R$, and the prime denotes differentiation with respect to $\phi$. 
The combination of Eqs. (\ref{148})--(\ref{150}) leads, finally,  to the equations of  energy-momentum conservation for  the various fluid components:
\bea
&&
\dot \r_r + 4 H \r_r=0,
\nonumber\\ &&
\dot \r_b + 3 H \r_b-{\psi'\over 2} \fpu \, \r_b =0,
\nonumber \\ &&
\dot \r_m + 3 H \r_m-{\psi'\over 2} \fpu\, \r_m -{\sg_m\over 2} \fpu=0,
\nonumber \\ &&
\dot \r_\phi + 3 H (\r_\phi+p_\phi)+{1\over 2} \fpu\left[ \psi' \left(\r_b+\r_m\right) +\sg_m \right]=0
\label{151} 
\eea
(the last equation is simply the dilaton equation (\ref{150}), rewritten in fluido-dynamical form). 

Let us now concentrate on the coupled dark-matter/dilaton system, and note that there are two types of interactions between these two cosmic sources: a first one, specific to the particular type of dark matter field, generated by the ``intrinsic" dilaton charge $\sg_m$; and a second one, more ``universal", generated by the standard dilaton coupling to the trace of the stress tensor, and associated to the $\psi'$ terms of the above equations. Both types of coupling are renormalized by the loop corrections, but with  opposite effect according to the asymptotic limits of Eq. (\ref{145}). In fact, the  dilaton charge tends to grow, and to reach a constant asymptotic value  as $\phi \ra +\infty$. The coupling parameter $\psi'$, on the contrary, tends to be exponentially suppressed as
\beq
 \psi'= - \left(\ln Z_R\right)' \ra {b_1 \e^{-\phi}\over c_1^2}, 
~~~~~~~~~~~~~~~~~~
\phi \ra +\infty.
\label{152}
\eeq
As a consequence, after the transition to the matter-dominated phase, 
the Universe may enter two different types of dynamical regimes \cite{64}.

$1)$ If the  dark-matter charge $\sg_m$ is still negligible at the beginning of the matter-dominated phase (as well as the dilaton potential, expected to become important only near the present epoch), then the Universe enters the so-called {\em ``dragging regime"}, in which $\r_m$ is coupled to $\phi$ through the  $\psi'$  terms of Eqs. (\ref{152}), and the evolution of the (still subdominant)  dilaton kinetic energy $\r_\phi$ is ``dragged" by $\r_m$. 

The cosmic evolution, during this regime, can be analytically described (in an approximate way) by noting that the loop factor $k(\phi)$ goes to a constant at late enough time scales,
\beq
k(\phi) \ra k_0= \sqrt{2} \,{c_2\over c_1}, ~~~~~~~~~~~~~~~~
\phi \ra +\infty, 
\label{153}
\eeq
according to Eqs. (\ref{148a}) and (\ref{145}). Introducing the canonical   variable $\wh \phi= k_0 \phi$ (see the action (\ref{147})), and neglecting the subdominant contributions of $\r_r$ and $\r_b$, we can then rewrite the coupled equations (\ref{150}), (\ref{151}), for the dragging regime, as follows:
\bea
&&
\ddot {\hf} + 3H\hfd +{\ep\over 2} \r_m=0,
\label{154}\\ &&
\dot \r_m +3H\r_m-{\ep\over 2} \r_m \hfd=0,
\label{155}
\eea
where $\ep= \psi'/k_0\simeq \e^{-\phi}/(\sqrt 2 c_1 c_2) \ll1$ is the effective coupling parameter. Neglecting the time dependence of $\ep$ with respect to that of $H$ and $\hfd$ (for small enough time intervals), we find that the system of equations (\ref{148}), (\ref{154}),  is satisfied by 
\beq
\hfd \simeq-2 \ep H. 
\label{156}
\eeq
Thus, from Eq. (\ref{155}), 
\bea
&&
\r_m \sim a^{-(3+\ep^2)} \sim H^2 \sim \hfd^2 \sim \r_\phi,
\nonumber\\ &&
a \sim t^{2/(3+\ep^2)}.
\label{157}
\eea
During this phase the dark-matter and the (kinetic) dilaton dark-energy densities are characterized by the same time-evolution, which slightly deviates from the standard behavior of a dust-dominated Universe ($\r \sim a^{-3}$, $a \sim t^{2/3}$). The kinematics, however, remains decelerated (as $\ep \ll1$). 

$2)$ A second, possibly accelerated, {\em ``freezing regime"} is eventually reached in the limit in which the dilaton potential comes into play, and the coupling induced by  the intrinsic charge density $\sg_m$ becomes dominant with respect to the exponentially suppressed coupling due to $\psi'$.

Using again the canonical variable $\hf$, assuming that $\sg_m= q(\phi) \r_m$ (for a homogeneous fluid), and considering the asymptotic limits $q(\phi) \ra q_0$, $V= V_0 \exp (-\phi)$ of Eq. (\ref{145}), we can rewrite the coupled dilaton-dark matter equations (\ref{151}), for the freezing regime, as follows: 
\bea
&&
\dot \r_m +3H\r_m-{q_0\over 2k_0} \r_m \hfd=0, 
\nonumber\\ &&
\dot \r_\phi +6H\r_k+{q_0\over 2k_0} \r_m \hfd=0.
\label{158}
\eea
We have defined the kinetic and potential energy densities, $\r_k$ and $\r_V$, respectively as 
\beq
\r_k= {\hfd^2\over 2}, ~~~~~~~  
\r_V= V(\wh \phi)=  V_0 \e^{-\hf/k_0},  ~~~~~~~~
\r_\phi= \r_k+ \r_V.
\label{159}
\eeq
The system of equations (\ref{158}), (\ref{148}) (with $\r_r=\r_b=0$) can be solved by a late-time configuration in which $\r_m$, $\r_\phi$, $V$ and $H^2$ scale in time in the same way, so that the critical fractions of dark-matter density, $\Om_m=\r_m/6H^2$, dilaton kinetic energy, $\Om_k=\r_k/6H^2$, and potential energy, $\Om_V=V/6H^2$, are separately frozen at constant values determined by $k_0$ and $q_0$ only (i.e. by the parameters $c_1$, $c_2$, $q_0$ of the asymptotic expansion (\ref{145})). A simple analysis gives \cite{64}
\bea
&&
\Om_k= {3k_0^2\over (q_0+2)^2}, ~~~~~~~~~~~~~
\Om_V={3k_0^2+q_0(q_0+2)\over (q_0+2)^2},
\nonumber\\ &&
\Om_\phi= \Om_k+\Om_V, ~~~~~~~~~~~~~~
\Om_m=1-\Om_\phi,
\label{160}
\eea
where $k_0$ is given by Eq. (\ref{153}) (see also \cite{26} for a detailed computation). 

In this asymptotic state the Universe is thus dominated by a fixed mixture of dark-matter and dilaton (kinetic plus potential) energy density. The dilaton fluid has equation of state
\beq
w={p_\phi\over \r_\phi}= {\Om_k-\Om_V\over \Om_k+\Om_V}=-
{q_0(q_0+2)\over 6 k_0^2 + q_0(q_0+2)},
\label{161}
\eeq
and can play the role of the dark energy fluid responsible for the observed cosmic acceleration, provided $q_0>1$.

In fact, by rewriting the Einstein equations (\ref{148}) for $\dot H$ in the form 
\beq
1+{2 \dot H\over 3 H^2}= \Om_V-\Om_k,
\label{162}
\eeq
we obtain 
\beq
{\ddot a\over a H^2}=1+ {\dot H\over H^2}= {3\over 2}\left(\Om_V-\Om_k\right)-{1\over 2}={q_0-1\over q_0+2}.
\label{163}
\eeq
The expansion is accelerated ($\ddot a >0$) for $q_0>1$ or $q_0<-2$. The second case (corresponding to an acceleration of superinflationary type, with $\dot H>0$) is to be excluded, however, in our context, as it would imply $\Om_m<0$ according to Eqs. (\ref{160}). Thus, acceleration is only possible for $q_0>1$. The explicit form of this asymptotic solution can be finally obtained through the integration of Eq. (\ref{163}), which gives
\beq
a \sim t^{(q_0+2)/3}, ~~~~~~~~~~~~
H \sim a^{-3/(q_0+2)},
\label{164}
\eeq
from which
\beq
\r_m  \sim H^2 \sim {\hfd^2\over 2} \sim  V_0\, \e^{-\hf/k_0} \sim a^{-6/(q_0+2)}.
\label{165}
\eeq

To illustrate the smooth background evolution from the initial radiation phase to the intermediate dragging phase, and to the final freezing regime, we shall conclude this subsection by presenting the results of an exact numerical integration of the string cosmology equations (\ref{148})--(\ref{151}). 
For our illustrative purpose we will assume that $Z_R$ and $Z_\phi$ are given by the expansion (\ref{145}) truncated to first order in $\exp(-\phi)$, with $b_1=b_2=1$, $c_1^2=100$ and $c_2^2=30$. We will adopt the model of dilaton coupling already used in \cite{64}, parametrized by the time-dependent charge
\beq
q(\phi)=q_0\,{\e^{q_0 \phi}\over c^2 + \e^{q_0\phi}},
\label{166}
\eeq
with $c^2=150$ and $q_0=2.5$. We will also use the E-frame potential (\ref{144}), with $\b_1=0.1$, $\b_2=0.2$, and $c_1^2 m_V= 10^{-3}H_{\rm eq}$. The last choice, which implies $m_V \sim H_0$, is crucial to obtain a realistic scenario in which the asymptotic accelerated regime starts at a phenomenologically acceptable epoch (see \cite{64,26} for a discussion of the mass scale of the non-perturbative dilaton potential, and of the degree of fine-tuning possibly required for realistic cosmological applications). Finally, we will integrate our equations imposing the initial conditions $\r_{\phi}(t_i)=\r_{r}(t_i)$, $\r_{m}(t_i)=10^{-20} \r_{r}(t_i)$, $\r_{b}(t_i)=7 \times 10^{-21}\r_{r}(t_i)$, $\phi (t_i)=-2$, at the initial scale $H (t_i)=10^{40}H_{\rm eq}$. 

\begin{figure}
\centering
\includegraphics[height=6cm]{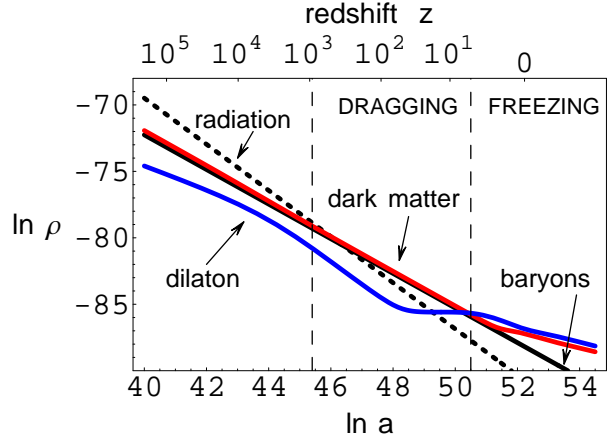}
\caption{Late-time evolution (on a logarithmic scale) of the various components of the cosmic energy density. The plots are the result of a numerical integration of Eqs. (\ref{148})--(\ref{151}).}
\label{f9}      
\end{figure}

The obtained scaling evolution $\r=\r(a)$ is illustrated in Fig. \ref{f9} for the various cosmic components. We can note that, at large enough times, baryons (full line) and radiation (dotted line) are fully decoupled from the dilaton, and obey the standard scaling behavior ($\r_r\sim a^{-4}$, $\r_b \sim a^{-3}$). The late-time dark-matter evolution, on the contrary, is closely tied to the dilaton evolution, and the ratio of their energy densities becomes asymptotically frozen at a constant. With the particular numerical values used in this example we obtain an asymptotic configuration characterized by $\Om_\phi\simeq 0.73$ and $\Om_m \simeq 0.27$, with a dark-energy equation of state $w \simeq -0.76$. 

\subsection{Non-local coupling and pressure backreaction}
\label{sec33}

Another interesting asymptotic configuration can be obtained in the case in which the dilaton is non-locally coupled to the dark matter components, as discussed in subsection \ref{sec12}. In that case, the fractions $\Om_m$ and $\Om_\phi$ may also become frozen at constant asymptotic values, but the background evolution turns out to be decelerated for any values of $q_0$, as the dark matter develops an effective pressure which tends to compensate the accelerating action of the dilaton potential. This  effect is new, and will be illustrated in some details in this subsection. 

We start assuming that, in the matter part of the action (\ref{114}),  the dilaton is non-locally coupled to the sources through the variable $\xi(\phi)$, as in the action (\ref{28}). However, differently from the action (\ref{28}), we will assume (for simplicity) that the dilaton potential is local, $V=V(\phi)$: the gravi-dilaton part of the action is thus identical to that of Eq. (\ref{114}), and our model is described by
\bea
&&
S=- {1\over 2\ls^2} \int \d^4 x \sqrt{-g} \left [Z_R(\phi) R+ Z_\phi(\phi) (\na \phi)^2 +V(\phi)\right]
\nonumber \\ && ~~~~~~~
+\int \d^4 x \sqrt{-g}\, \cl(\e^{-\xi}).
\label{167}
\eea
Let us vary this action with respect to $g$ and $\phi$, and evaluate the resulting (general covariant) field equations in the limit of a homogeneous, isotropic, spatially flat background, using the results of Subsection \ref{sec12}. We obtain a set of equations similar to Eqs. (\ref{45})--(\ref{47}) for what concerns the dilaton charge density $\sg(\fb)$, but different for the potential (which now is local), and for the presence of the loop corrections $Z_R$, $Z_\phi$. 

Let us finally transform the equations in the E-frame (using the rescaling (\ref{146})), and consider the asymptotic limit in which $\r_r$, $\r_b$ are negligible, and the dark matter is coupled to the dilaton only through its intrinsic dilaton charge (namely, the limit in which $\psi' \simeq 0$). The resulting equations (omitting the tilde, in units $2 \lp^2=1$) can be written as
\bea
&&
6H^2 = \r_m+\r_\phi,
\label{168}\\ &&
4 \dot H+ 6H^2=  - p_\phi- {\sg_m\over 2},
\label{169}\\ &&
\dot \r_m + 3 H\left( \r_m+{\sg_m\over 2}\right) -{\sg_m\over 2} \fpu=0,
\label{170} \\ &&
\dot \r_\phi + 3 H (\r_\phi+p_\phi) +{\sg_m\over 2} \fpu =0,
\label{171} 
\eea
with $\r_\phi$ and $p_\phi$ defined by Eq. (\ref{149}), as before. A comparison with the asymptotic limit of Eqs. (\ref{148}), (\ref{151}), shows that the genuinely new effect of the non-local interactions is the appearance of an  effective pressure
term $\sg_m/2$ for the dark matter component. Indeed, the new terms present in Eqs. (\ref{169}), (\ref{170}), can also be obtained from the standard Einstein equations through the shift $p_m \ra p_m+ 
\sg_m/2$.

We are now in the position of asking whether or not this modification 
(of non-local origin) may change the results of the previous subsection, in particular those concerning the asymptotic freezing configuration. We shall consider, to this purpose, the limit in which $\sg_m \ra q_0\r_m$ and $V=V_0 \exp(-\phi)$, using  the canonical variable $\hf$ as in the previous computations. 

Let us look for solutions of Eqs. (\ref{168})--(\ref{171}) by requiring for $\r_m$, $\r_k$ and $\r_V$ the same scaling behavior, and thus imposing, as  a first condition, that  
\beq
{\dot \r_m\over \r_m}= {\dot \r_\phi\over \r_\phi}.
\label{172}
\eeq
Using Eqs. (\ref{170}), (\ref{171}) for $\r_m$ and $\r_\phi$, and the Einstein equation (\ref{168}), we obtain
\beq
{\hfd\over H}= {6 k\over q_0}\left[\overline \Om_V\left(1+{q_0\over 2}\right)-\overline\Om_k\left(1-{q_0\over 2}\right)\right]. 
\label{173}
\eeq
We are denoting with a bar the fractions of critical energies for the new configuration associated to the non-local equations, to distinguish it from the ``local" freezing solution of Eq. (\ref{160}). We also impose, as a second condition,  that
\beq
{\dot \r_m\over \r_m}= {\dot \r_V\over \r_V}.
\label{174}
\eeq
The definition (\ref{159}) of $\r_V$, together with Eq. (\ref{170}), gives then
\beq
{\hfd\over H}=3 k_0,
\label{175}
\eeq
which, combined with eq. (\ref{173}), leads to
\beq
\overline\Om_V=\overline \Om_k \,{2-q_0\over 2+q_0}+ {q_0\over q_0+2}.
\label{176}
\eeq
From the definition of $\overline \Om_k$ and Eq. (\ref{175}), on the other hand, we have
\beq
\overline \Om_k = {\hfd \over12\, H^2}= { 3 \over 4} \,k_0^2.
\label{177}
\eeq
The insertion of this result into Eq. (\ref{176}) finally gives
\beq
\overline \Om_V={3 k_0^2(2-q_0)+4 q_0\over 4(2+q_0)}.
\label{178}
\eeq

The combination of $\overline \Om_k$ and $\overline \Om_V$ provides now the values of $\overline \Om_\phi$, $\overline \Om_m$, and the equation of state $\overline w$, according to the definitions (\ref{160}), (\ref{161}). As we are interested in the kinematical properties of the solution we shall compute, in particular, the acceleration parameter $\ddot a/(aH^2)$: dividing by $6H^2$ the modified equation (\ref{169}) we obtain
\beq
{\dot H\over  H^2}= {3\over 2}\left(\overline\Om_V-\overline\Om_k\right)-{3\over 4} q_0 \,\overline\Om_m -{3\over 2},
\label{179}
\eeq
from which
\beq
{\ddot a\over a H^2}\equiv1+ {\dot H\over H^2}= -{1\over 2},
\label{180}
\eeq
quite independently of the values of $k_0$ and $q_0$! The integration of $\dot H$ finally provides $a \sim t^{2/3}$ and $H^2 \sim \r \sim a^{-3}$, as in the standard phase of dark-matter dominated evolution. 

The considered model of non-local coupling is thus associated to an asymptotic freezing phase which is decelerated, and in which the dilaton energy density has the same dynamical behavior of a dust fluid, $\r_\phi \sim \r_m \sim a^{-3}$, in spite of a pressure which is nonvanishing, in general:
\beq
\overline w= {q_q\over 2}\, {3 k_0^2-2\over 3 k_0^2+q_0}.
\label{181}
\eeq
This result can be understood by noting Eqs. (\ref{179}) and (\ref{180}) together imply
\beq
\overline\Om_k-\overline\Om_V+{q_0\over 2}\, \overline\Om_m
\equiv {1\over 6H^2} \left(p_\phi+{\sg_m\over 2}\right)=0,
\label{182}
\eeq
namely a zero  total pressure for the coupled dilaton-dark matter fluid (see Eq. (\ref{169}). The dark matter pressure associated to the  non-local effects  thus generates a backreaction which exactly compensates -- at least in this model -- the dilaton pressure, leading the system to restore, asymptotically, the standard dust matter configuration. 

\subsection{Main differences from uncoupled models}
\label{sec34}

Let us come back to the class of models in which the dilaton is locally coupled to the dark matter components, as discussed in subsection \ref{sec32}. If we identify the accelerated freezing phase with our present cosmological phase, and thus the energy density of the dilaton field with the ``dark-energy" density responsible for the present cosmic acceleration, we are lead to a dilaton model of dark energy  which is substantially different from the conventional models of quintessence \cite{67} based on a rolling scalar field, uncoupled to dark matter.

A first, important (conceptual) difference concerns the mentioned problem of the cosmic coincidence. In the considered class of dilaton models this problem, if not solved, is at least relaxed: in fact, the 
dark-energy and dark-matter densities are of the same order not only today but also in the future (forever), and also in the past for a significantly amount of time, depending on the beginning of the freezing epoch (see below). 

A second, more phenomenological difference concerns the scaling behavior of the baryonic and dark-matter components of the dust fluid during the freezing epoch. Because of the coupling to the dilaton, the dilution in time of the dark-matter density $\r_m$ is slower than the standard baryon dilution, $\r_b \sim a^{-3}$: in particular, the ratio $\r_b/\r_m$ decreases in time as 
\beq
{\r_b\over \r_m}\sim a^{-3 q_0/(2+q_0)}
\label{183}
\eeq
(see Eq. (\ref{165})). This could explain why the present fraction of baryons is small $(\sim 10^{-2}$) in critical units -- provided the accelerated epoch has an early enough beginning. Direct/indirect measurements of the past value of the ratio $\r_b/\r_m$, compared with its present value, could provide unambiguous tests of this class of models. 

Finally, concerning the beginning of the accelerated epoch, it is important to stress that in dilaton models the acceleration can start much earlier than in models of uncoupled quintessence \cite{68,69}. 

For a simple illustration of this point we may consider a model in which, during the accelerated regime, there are two types of sources with different dynamical behavior: $i)$ an {\em uncoupled} component $\r_u$, with pressure $p_u=0$ and scaling behavior $\r_u\sim a^{-3}$ (represented by baryons and,  possibly, by a fraction of non-baryonic dark matter uncoupled to the dilaton); $ii)$ a {\em coupled} component $\r_c$, with pressure $p_c=w \r_c$, and a slower scaling behavior $\r_c\sim a^{-3(1+w)}$ (represented by the dilaton and by the   fraction of  dark matter coupled to the dilaton). Thus, even if today $\r_c$ dominates, and drives an accelerated evolution, at ealy enough times the Universe was dominated by $\r_u$, and decelerated. From the Einstein equations 
\bea
&&
6H^2= \r_u+\r_c, 
\nonumber \\ &&
4 \dot H +6H^2 =-p_c=-w\r_c, 
\label{184}
\eea
we obtain that the acceleration switches off at the scale $a_{acc}$ such that 
\bea
&&
\left(\ddot a \over a H^2 \right)_{\rm acc}= 1 + \left(\dot H \over H^2\right)_{\rm acc}
=-{1\over 2} \left[\Om_u-(1+3w)(\Om_u-1)\right]_{\rm acc}=0,
\nonumber \\
&& 
\label{185}
\eea
where $\Om_u=\r_u/6H^2$, $\Om_c=1-\Om_u$. In terms of the present values $\Om_u^0$, $\Om_c^0$ of these fractions the above condition becomes
\beq
\Om_u^0\left(a_{\rm acc}\over a_0\right)^{-3}=(1+3w)\left(\Om_u^0-1\right) \left(a_{\rm acc}\over a_0\right)^{-3(1+w)},
\label{186}
\eeq
and fixes the beginning of the acceleration at the redshift scale $z_{acc}$ such that
\beq
z_{\rm acc}\equiv{a_0\over a_{\rm acc}}-1= \left[(1+3w)\left(\Om_u^0-1\over \Om_u^0\right)\right]^{-1/3w}-1.
\label{187}
\eeq

\begin{figure}
\centering
\includegraphics[height=3.5cm]{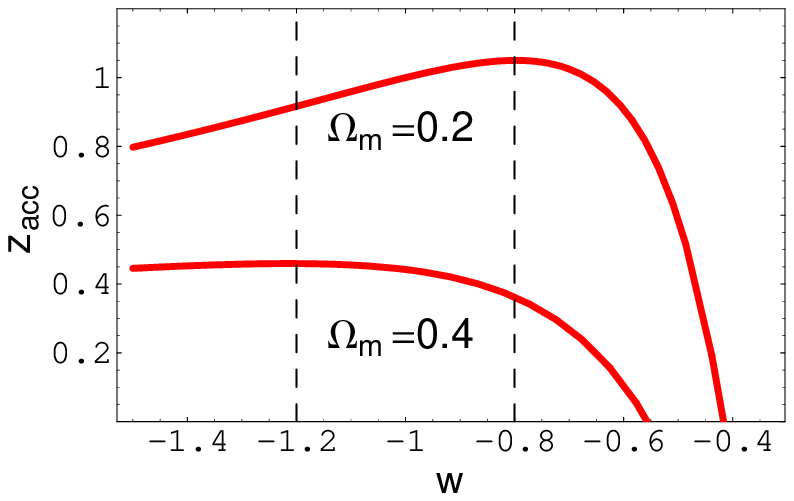}~
\includegraphics[height=3.5cm]{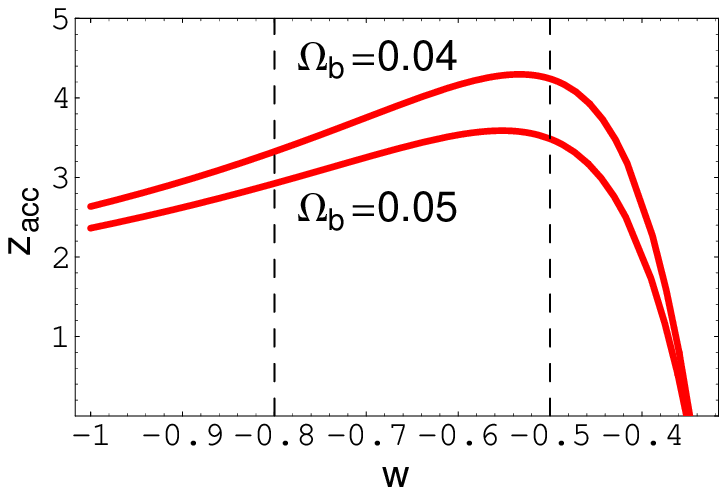}
\caption{Beginning of the accelerated epoch for dark-nergy models with uncoupled (left) and fully coupled (right) dark matter, according to the observations of type Ia supernovae in the SNLS dataset. The plotted curves are obtained from Eq. (\ref{187}), for constant values of the present fraction of the uncoupled dust matter $\Om_u^0$.}
\label{f10}      
\end{figure}

Consider now a model of uncoupled quintessence, in which the uncoupled component corresponds to the totality of the dark-matter fluid (plus subdominant contributions), i.e. $\Om_u^0= \Om_m^0$. Using the recent analysis of the SNLS Collaboration \cite{61a}, based on present observations of supernovae Ia and large-scale structure, one finds $0.2 \leq \Om_m^0 \leq 0.4$, and $-1.2 \leq w\leq 0.8$. One then obtains, from Eq. (\ref{187}), $0.4 \laq z_{acc} \laq 1$ (see Fig. \ref{f10}, left panel). 

If we consider instead a model of dilaton dark energy, then the uncoupled component may range from the baryon component 
$\Om_b$ to some fraction of the dark matter component $\Om_m$. In the ``maximally coupled" version of the model, in which $\Om_u^0=\Om_b^0$, we can re-apply the supernovae results of SNLS with $\Om_b^0 \simeq 0.04-0.05$, to obtain $ w \simeq -0.65 \pm 0.15$. Eq. (\ref{187}) then implies  \cite{69} $z_{acc} \simeq 3-4$ (see Fig. \ref{10}, right panel). 

Thus, dilaton models of dark energy are compatible with a beginning of the cosmic accelerations at epochs much earlier than those suggested by other models, according to the most recent supernovae data. The extension back in time of the accelerated regime might have a significant impact on the dilution of baryons, according to Eq. (\ref{183}). Finally, strongly coupled models tend to be compatible with a ``less negative" parameter $w$  (see Fig. \ref{10}), thus alleviating the need for ``phantom" dark energy \cite{70} with ``supernegative" ($w<-1$) equation of state. 

\section*{Acknowledgements}

I am very grateful to all friends and collaborators contributing to the results reported in this paper. First of all I would like to thank  Gabriele Veneziano for many years of collaborations, friendships and support. In addition, I am grateful to Valerio Bozza, Massimo Giovannini and Jnan Maharana for their collaboration on the results presented in the first lecture; to  Nicola Bonasia, Eugenio Coccia and Carlo Ungarelli for their collaboration on the results presented in the second lecture; to Luca  Amendola, Federico Piazza and Carlo Ungarelli for their collaboration on the results presented in the third lecture. 

\begin{appendix}
\renewcommand{\theequation}{A.\arabic{equation}}
\setcounter{equation}{0}
\section*{Appendix A}

In this Appendix we present  a detailed derivation of the equations of motion (\ref{29}), (\ref{34}), starting from the action (\ref{28}) which includes non-local dilaton interactions. 

The functional derivation of the action with respect to the metric $g^{\mu\nu}(x)$ contains, besides the standard contributions leading to Eq. (\ref{3}), the new non-local contributions $V_{\mu\nu}(x)$ and $L_{\mu\nu}(x)$, and can be written as follows,
\bea
{\da S\over \da g^{\mu\nu}(x)}= &-&{\left(\rg\, \e^{-\phi} \right)_x\over 2 \ls^{d-1}}  
\left[ G_{\mu\nu}+ \na_\mu\na_\nu \phi +{1\over 2} g_{\mu\nu}\left(
\na \phi^2 -2 \na^2 \phi -V\right) \right]_x
\nonumber \\ &+&
{1\over 2} \rg\, T_{\mu\nu}(x) +V_{\mu\nu}(x)+L_{\mu\nu}(x),
\label{A1}
\eea
where 
\bea
&&
V_{\mu\nu}(x)= -{1\over 2 \ls^{d-1}}\int \d^{d+1} x' \left( \rg \, \e^{-\phi} V'\right)_{x'} 
{\da\over \da g^{\mu\nu}(x)} \e^{-\xi(x')}, 
\label{A2}\\ &&
L_{\mu\nu}(x)= \int \d^{d+1} x' \left( \rg \cl'\right)_{x'} 
{\da\over \da g^{\mu\nu}(x)} \e^{-\xi(x')}
\label{A3}
\eea
($V'$ and $\cl'$ are defined by Eq. (\ref{33})). We need now to compute the functional derivation of $\exp(-\xi)$. Using the definition (\ref{25}) we obtain  
\bea
&&
{\da\over \da g^{\mu\nu}(x)} \e^{-\xi(x')}=-{1\over \ls^d} \int \d^{d+1}y \,\da^{d+1}(x-y) \da(\phi_{x'}-\phi_y) \, \e^{-\phi_y}
\nonumber \\ &&
~~~~~~~~~~~~
\left[-{1\over 2} \left(\rg \, g_{\mu\nu} \rp\right) +{1\over 2} \rg \rp ~{\pa_\mu \phi \pa_\nu \phi \over (\na \phi)^2} \right]_y
\nonumber \\ && ~~~~~~~~~~~~
=-{1\over 2 \ls^d} \left( \ga_{\mu\nu} \rg \, \rp\, \e^{-\phi}\right)_x
\da(\phi_{x'}-\phi_x),
\label{A4}
\eea
where $\ga_{\mu\nu}$ is defined in Eq. (\ref{30}). Thus:
\bea
&&
V_{\mu\nu}= {1\over 2 \ls^{d-1}}\, \rg \, \e^{-2\phi} {1\over 2} \ga_{\mu\nu} \rp \, I_V,
\label{A5} \\ &&
L_{\mu\nu}= - \rg\, \e^{-\phi} {1\over 2} \ga_{\mu\nu} \rp \, I_m,
\label{A6}
\eea
where $I_V$ and $I_m$ are defined in Eqs. (\ref{31}), (\ref{32}). Inserting these results in Eq. (\ref{A1}), multiplying by $(-2 \ls^{d-1}) \exp(-\phi)/ \rg$, and imposing the condition of zero functional derivative, one is finally lead to Eq. (\ref{29}). 

Let us now consider the functional derivative with respect to $\phi(x)$. Separating the local and non-local terms, as before,  we obtain
\bea
{\da S\over \da \phi(x)} &=&{\left(\rg\, \e^{-\phi} \right)_x\over 2 \ls^{d-1}}  
\left[ R+ 2 \na^2 \phi-(\na \phi)^2 +V \right]_x
\nonumber \\ &+&
A(x)+ B(x),
\label{A7}
\eea
where 
\bea
&&
A(x)= -{1\over 2 \ls^{d-1}}\int \d^{d+1} x' \left( \rg \, \e^{-\phi} V'\right)_{x'} 
{\da\over \da \phi(x)} \e^{-\xi(x')}, 
\label{A8}\\ &&
B(x)= \int \d^{d+1} x' \left( \rg \cl'\right)_{x'} 
{\da\over \da \phi(x)} \e^{-\xi(x')}.
\label{A9}
\eea
The functional derivative of the variable (\ref{25}) leads to 
\bea
&& 
{\da\over \da \phi(x)} \e^{-\xi(x')}= 
\nonumber \\ &&=
{1\over \ls^d} \int \d^{d+1}y
\Bigg\{ -\left(\rg \, \e^{-\phi} \rp\right)_y  \da(\phi_{x'}-\phi_y)
\da^{d+1}(x-y)
\nonumber \\ &&+
\left(\rg \, \e^{-\phi} \rp\right)_y \da'(\phi_{x'}-\phi_y)
\left[ \da^{d+1}(x-x')-\da^{d+1}(x-y)\right]
\nonumber \\ &&-
 \pa_\mu \left[ {\rg\, \e^{-\phi} \ep\, \pa^\mu \phi \over \rp} 
\da(\phi_{x'}-\phi_y)\right]_y \da^{d+1}(x-y)
\Bigg\}, 
\label{a10}
\eea
where $\pa_\mu= \pa/\pa y^\mu$, and $\da'$ denotes the derivative of the delta function with respect to its argument. 

The first term of this integral exactly cancels the term containing $\pa_\mu \e^{-\phi}$ in the last part of the integral; also, the third term exactly cancels  the term containing $\pa_\mu [\da(\phi_{x'}-\phi_y)]$ in the last part of the integral. Thus, we are left with:
\bea
{\da\over \da \phi(x)} \e^{-\xi(x')}&=&{\da^{d+1} (x-x')\over \ls^d} \int \d^{d+1}y \left(\rg \e^{-\phi} \rp \right)_y \da' (\phi_{x'}-\phi_y) 
\nonumber \\ &-& 
\ep \,{\e^{-\phi} \over \ls^d} \da(\phi_{x'}-\phi_x)\pa_\mu  \left( \rg \,\pa^\mu \phi\over \rp\right)_x.
\label{A11}
\eea
The second term on the right-hand side of the above equation can be conveniently rewritten as 
\bea 
&&
-\ep {\e^{-\phi}\over \ls^d} \da(\phi_{x'}-\phi_x)\rg\, \na_\mu  \left( \pa^\mu \phi\over \rp\right)_x=
\nonumber \\ &&
=- \ep {\e^{-\phi}\over \ls^d} \da(\phi_{x'}-\phi_x){\rg \over \rp} \,\ga_{\mu\nu} \na^\mu \na^\nu \phi. 
\label{A12}
\eea
For the first term containing $\da'$ we can exploit the properties of the delta function, and the identities
\beq
\d y_0= {1\over \dot \phi_y}{\d \phi_y}, ~~~~~~~~~~
{\d \over \d \phi_y}= {1\over \dot \phi_y} {\d \over \d y_0},
\label{A13}
\eeq
to obtain 
\bea
&&
\ls^{-d} \int \d^{d}y \,{\d \phi_y\over \dot \phi_y} 
\left(\rg\,\e^{-\phi} \rp \right)_y \da' (\phi_{x}-\phi_y)
\nonumber \\ &&
= \ls^{-d} \int \d^{d}y \,{\d \phi_y\over \dot \phi_y} \,{\d \over \d y_0} 
\left(\rg\,\e^{-\phi} \rp \over \fpu \right)_y \da (\phi_{x}-\phi_y)
\nonumber \\ &&
=-{e^{-\xi(x)}}+ 
\ls^{-d} \e^{-\phi(x)}\int \d^{d}y \,{\d \phi_y\over \dot \phi_y} \,\left(\rg\,\rp  \right)_y \da' (\phi_{x}-\phi_y)
\nonumber \\ &&
=-{e^{-\xi(x)}} + e^{-\phi(x)} J(x),
\label{A14}
\eea
where $J$ is the integral defined in Eq. (\ref{35}). Inserting the results (\ref{A12}), (\ref{A14}) in Eq. (\ref{A11}), using the definitions of $A$ and $B$, and integrating over $x'$, we finally obtain
\bea
A(x)+B(x)&=&  \left({\rg \, \e^{-\phi}\over 2 \ls^{d-1}}\, V' - 
\rg \,\cl'\right)_x \left({e^{-\xi}} - e^{-\phi} J\right)_x
\nonumber \\ &+&
 \ep \, {\rg \,\e^{-\phi}\over \rp} \ga_{\mu\nu} \na^\mu \na^\nu \phi 
\left({\e^{-\phi}\over 2\ls^{d-1}}\, I_V- I_m\right)_x.
 \label{A15}
 \eea
Summing this contribution to the other terms of Eq. (\ref{7}), multiplying by $(2 \ls^{d-1} \exp \phi /\rg)$, and imposing the vanishing of the functional derivative, we are lead to the equation of motion (\ref{34}) for the dilaton.

\end{appendix}

\index{paragraph}


\printindex
\end{document}